\documentclass[
manuscript=article,  
layout=publish,  
year=2023,
volume= ,
]{joas}

\received {1 April 2022}
\revised  {1 May 2022}
\accepted {10 May 2022}
\published{20 May 2022}
\editor{Editor Name}
\reviewers{First Reviewer, Second Reviewer, Third Reviewer}

\usepackage[english]{babel}

\usepackage[utf8]{inputenc}
\usepackage[T1]{fontenc}
\usepackage{amsmath}
\usepackage{amsfonts}
\DeclareMathAlphabet{\mathsfit}{T1}{\sfdefault}{\mddefault}{\sldefault}
\SetMathAlphabet{\mathsfit}{bold}{T1}{\sfdefault}{\bfdefault}{\sldefault}

\usepackage[version=4]{mhchem}
\usepackage{stmaryrd}
\usepackage{bbold}
\usepackage{soulutf8}
\usepackage{graphicx}
\usepackage[export]{adjustbox}
\graphicspath{ {./images/} }

\usepackage{subcaption}

\usepackage{hyperref}

\usepackage{listings}
\usepackage{color}

\definecolor{mygreen}{rgb}{0,0.6,0}
\definecolor{mygray}{rgb}{0.5,0.5,0.5}
\definecolor{mymauve}{rgb}{0.58,0,0.82}

\usepackage[style=numeric,backend=biber,sorting=none]{biblatex} 
\title{Comparison of Neural FEM and Neural Operator Methods for applications in Solid Mechanics}

\author{Stefan Hildebrand}
\affiliation{TU Berlin, Berlin, Germany}
\email{stefan.hildebrand@tu-berlin.de}

\author{Sandra Klinge}
\affiliation{TU Berlin, Berlin, Germany}


\begin{document}
	
	\begin{abstract}
		Machine Learning methods belong to the group of most up-to-date approaches for solving partial differential equations. The current work investigates two classes, Neural FEM and Neural Operator Methods, for the use in elastostatics by means of numerical experiments. The Neural Operator methods require expensive training but then allow for solving multiple boundary value problems with the same Machine Learning model. Main differences between the two classes are the computational effort and accuracy. Especially the accuracy requires more research for practical applications.
	\end{abstract}

	\section{Introduction}	
	Induced by ever-rising compute power and successful applications in several domains, Artificial Intelligence (AI) systems and especially Machine Learning (ML) methods attract growing attention for advanced tasks in mechanical engineering \cite{Bock2019, KIRCHDOERFER2016, Shoghi2022}. This is supported by well-established and flexible Machine Learning frameworks like PyTorch \cite{PyTorch2019} and Tensorflow \cite{tensorflow2015-whitepaper}.
	
	One particular application of ML is the solution of Parameterized Partial Differential Equations (PPDE), which are traditionally solved by numerical discretization methods like Finite Element Method (FEM), Finite Difference Method (FDM), Finite Volume Method (FVM) or Boundary Element Method (BEM). Based on ML techniques, two new classes of methods arose, namely the Neural FEM and Neural Operator methods \cite{Li2020b}. The aim of the current work is to compare these two classes of methods for applications in solid body mechanics. Therefor, their common representatives are applied to case studies, where the well-established FEM can serve as a benchmark.

	The mathematical problem to solve with either method can be described as follows:
	
	Let an arbitrary Parameterized Partial Differential Equation be given on an open domain $B$ with piecewise smooth boundary $\Gamma$ in the form:
	
	\begin{align}
		\mathcal{N}[\boldsymbol{u}(\boldsymbol{y}) ; \boldsymbol{y}]=\mathbf{0} \quad \text{on} \,\, B, \quad \mathcal{B}[\boldsymbol{u}(\boldsymbol{y}) ; \boldsymbol{y}]=\mathbf{0} \quad \text{on} \,\, \Gamma \,\, ,
	\end{align}
	
	where $\mathcal{N}$ is a nonlinear operator on the domain $B$, $\mathcal{B}$ an operator on $\Gamma$ that determines the boundary conditions, and $\boldsymbol{u}(\boldsymbol{y})\in  \mathbb{R}^{d}$ the solutions of the PDE. All quantities are parameterized by $\boldsymbol{y} \in \mathbb{R}^n$. The mapping
	
	\begin{align}
		G: \quad B \cup \Gamma \,\, \times \,\, \mathbb{R}^n \rightarrow \mathbb{R}^{d}, \quad \left(\boldsymbol{X}, \boldsymbol{y}\right) \mapsto \boldsymbol{u}, \quad \boldsymbol{X} \in B \cup \Gamma, \quad n, d \in \mathbb{N}
	\end{align}
	 is called the solution operator of the PPDE.

	Neural FEM resembles a conventional FEM implementation.
	The artificial Neural Network (NN) approximates the solution function of a particular realization of the PPDE. Based on the conventional form of Physics-Informed Neural Networks (PINN, \cite{Raissi2019}), the Deep Energy Method (DEM) and competitive PINNs (cPINN) are proposed in \cite{Nguyen-Thanh2020, Zeng2022} . All these approaches are independent from a spatial discretization of the domain $B$ (grid-independent) and can realize high accuracies, but must be retrained for each new set of parameters. 
	
	In the case of Neural Operator methods, an NN is trained to behave like the solution operator of a PPDE. Then, the network can be applied to arbitrary combinations of parameters and boundary conditions, to solve Boundary Value Problems (BVP).
	These methods are particularly characterized by a discretization-independent error, allowing zero-shot super resolution (training on coarse grid, inference on fine grid). 
	In this work, the Deep Operator Network (DeepONet) \cite{Lu2021} and the Fourier Neural Operator (FNO) \cite{Li2020b} are studied as representatives of Neural Operator methods. 
	Typically, Neural Operator Methods require a large amount of training data,  which may need to be computed in a numerically expensive way \cite{Kollmannsberger2021}.
	
	Physics-Informed variants of Neural Operator methods address this drawback by incorporating knowledge on the underlying PDE as a regularizing mechanism in the loss function \cite{Raissi2019}. This can increase accuracy, generalizability, and data efficiency \cite{Wang2021b}. The present contribution investigates Physics-Informed DeepONet (PIDeepONet) \cite{Wang2021b} and Physics Informed Neural Operator (PINO) \cite{Li2022} as representatives of Physics-Informed Neural Operator methods.
	
	The paper is structured as follows. First, we give an insight in details of selected ML methods (Sections \ref{sec:3-1} and \ref{subsec:neural-operator-methods}). Then, we apply these methods to three example problems (Section \ref{sec:Application-problems}). Finally, we discuss the results in comparison to the reference solution from FEM and highlight necessary steps for a future use of these NN methods in elastostatics (Section \ref{sec:SumOut}).
	
	\section{Neural Network Nomenclature}
	Artificial Neural Networks are constructed as layers of neurons \cite{Abueidda2021, Bock2019}.
	Each neuron carries out a (typically nonlinear) activation function. In case of a Fully Connected Neural Network (FCNN), each neuron receives its input as linear transformation of all outputs of the neurons on the layer before. 
	The output $\boldsymbol{\mathcal{R}}^i$ of the $i^{th}$ layer is thus calculated by:
	\begin{align}
		\boldsymbol{W}^i &= \boldsymbol w^{i} \boldsymbol{\mathcal{R}}^{i-1} + \boldsymbol{b}^i \\
		\boldsymbol{\mathcal{R}}^i &= \boldsymbol{a}^i(\boldsymbol{\Theta}^i,  \boldsymbol{W}^i)
	\end{align}
	The weights $\boldsymbol w^{i}$ and biases $\boldsymbol{b}^i$ of the $i^{th}$ layer, together with the parameters $\boldsymbol{\Theta}^i$ of the activation functions $\boldsymbol{a}$, form the set of parameters $\theta$ of the Neural Network. Other parameters of the network architecture (like number of layers and layer widths) and the optimizer algorithm (e.g. step width) are called hyperparameters and have to be chosen by the NN user or an outer optimization strategy.
	
	All the layers between the input an output layer are called hidden layers. The number of (hidden) layers is referred to as the network's depth, whereas the number of neurons within a layer is called the width of the layer. 
	
	The whole network represents an arbitrary (continuous) mapping (Universal approximation theorem, \cite{Hornik1989}) $\mathcal{R}: \mathbb{R}^n \mapsto \mathbb{R}^m$ from the input to the output side, with $n$ and $m$ the input and output layer width, respectively.
	
	To make the network approximate a mapping $\tilde{\mathcal{R}}$ on a subset $D \subset \mathbb{R}^n$, the mapping is given indirectly by a set $T_{tr}$ of training tuples $t_{tr}^k \left(P^k, \tilde{\mathcal{R}}(P^k) \right), \,\, P^k \in D, \,\, k \in \mathbb{N}$, which together form the training data set. $P^k$ are the input samples, $\tilde{\mathcal{R}}(P^k)$ the corresponding target outputs. A set $T_{te}$ of testing tuples $t_{te}^l, \,\, l \in \mathbb{N}$  is required to check the quality of the approximation the network has learned so far. Usually, $T_{tr} \cap T_{te} = \emptyset$. In the application of a network, arbitrary sets of data within $D$ can be the input, but the exact output is usually unknown and only approximated by the net.
	
	Conventionally, the parameters of the network are adapted by an optimization algorithm like Adam \cite{Kingma2014}. This algorithm minimizes the empirical risk $F$ (alternatively called loss value) that is calculated as the output of a loss function $\mathcal{L}$. The latter is commonly defined as the discrepancy between the target outputs and the outputs of the network with the current parameters. A frequent choice for the loss function is Mean Square Error (MSE) for $N \in \mathbb{N}$ tuples $t^k$ 
	\begin{align}
		\mathrm{MSE} = \frac{1}{N} \sum_{k=1}^{N} \left(\tilde{\mathcal{R}}(P^k) - \mathcal{R}(P^k) \right)^2 \quad. \label{eq:MSE}
	\end{align}
	
	The optimization employs (partial) derivatives of the loss function w.r.t.\@ the parameters in the network. Machine learning networks like PyTorch therefor record all operations acting on a variable from input to output symbolically, so that fast, highly accurate derivation becomes possible. This feature is referred to as autograd \cite{Paszke2017}. Its use, however, is not limited to derivations w.r.t.\@ network parameters.

	\begin{figure}
		\centering
		\includegraphics[width=0.8\textwidth]{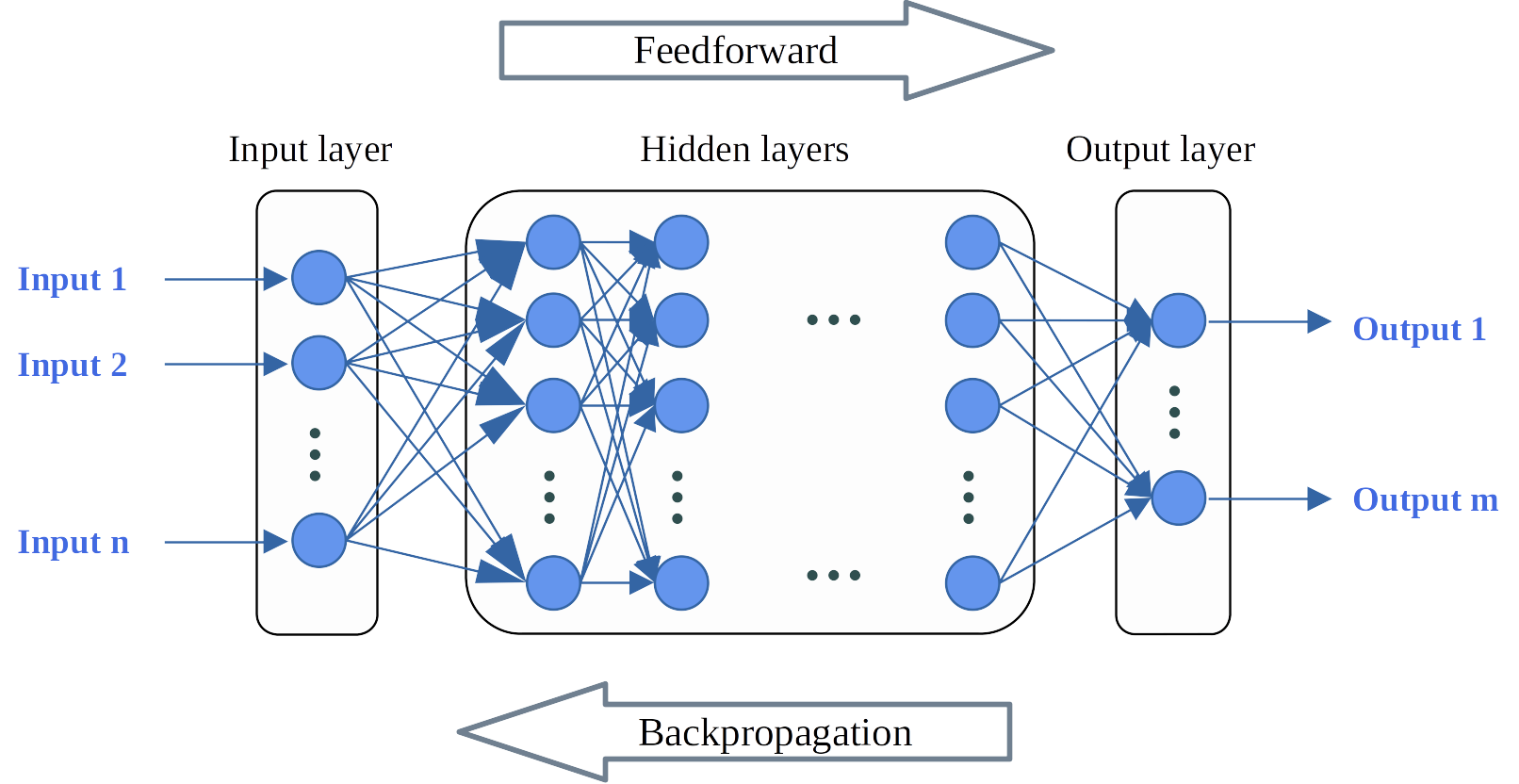}
		\caption[FCNN]{Information flow in a Fully Connected Neural Network (FCNN).}
		\label{fig:fcnn-pic}
	\end{figure}

	By default, the NN parameters are initialized randomly before the first optimizer step. This leads to varying results of effort and achieved accuracy during training.

	\section{Neural FEM} \label{sec:3-1}
	In this class of methods, the output of the NN is chosen as the unknown function of the PDE. 
	The loss function is then computed either from the residual of the PDE (classical Physics-informed Neural Networks (PINN) \cite{Raissi2019} and competitive PINN (cPINN) \cite{Zeng2022}), or from the potential energy if the minimum principle applies (Direct Energy Method (DEM) \cite{Nguyen-Thanh2020}, mixed DEM (mDEM) \cite{FuhgBouklas2022}), both incorporating the outputs of the NN. Hence, no training data set is needed.

	\subsection{Physics Informed Neural Networks (PINN)}
	The original PINN formulation uses an FCNN where the loss function is applied to the squared residual of the PDE on specified collocation points, \cite{Raissi2019}, 
	\begin{align}
		F^{\mathrm{PINN}, B} = \frac{1}{N_{f}} \sum_{i=1}^{N_{f}}\left(\mathcal{N}\left[\boldsymbol{u}_{\theta}\right]\left(\boldsymbol{X}_{i}^{f}\right)\right)^{2} \quad .
	\end{align}
	The boundary conditions are accounted for by an additional term in the form
	\begin{align}
		F^{\mathrm{PINN}, \Gamma} = \frac{\lambda_{b}}{N_{b}} \sum_{i=1}^{N_{b}}\left(\mathcal{B}\left[\boldsymbol{u}_{\theta}\right]\left(\boldsymbol{X}_{i}^{b}\right)\right)^{2},
	\end{align}
	where $\lambda_{b}$ is a hyperparameter that weighs the error proportions, since the propagated gradients can be of different magnitudes, thus driving the optimization procedure towards an incorrect solution \cite{Wang2021a}.
	The total empirical risk is then calculated by	 
 	\begin{align}
	 	F^{\mathrm{PINN}}\left(\boldsymbol{u}_{\theta}\right)=
	 	F^{\mathrm{PINN}, B} + F^{\mathrm{PINN}, \Gamma} \quad .
	 \end{align}

	According to \cite{Krishnapriyan2021}, the training of PINNs fails even on very simple problems, such as the 1D convection	or the reaction-diffusion equation.
	An analysis of the occurrence of comparable pathologies in elastostatic or elastodynamic contexts requires further research. In the survey at hand, they did not manifest.
	
	\subsubsection{Deep Collocation Method (DCM)} \label{subsubsec:DCM}
	The Deep Collocation Method (DCM) \cite{Abueidda2021} is a representative of the classical PINN, where the empirical risk is built up by the squared residual at random collocation points. The constraints are typically accounted for by additional penalty terms but not enforced by a transformation of the output data.

	\subsection{The Deep Energy Method (DEM)}
	The DEM was originally introduced to calculate finite deformation hyperelasticity \cite{Nguyen-Thanh2020}.
	This method as well as methods derived from it require only first derivatives to compute the loss function, thus reducing the numerical complexity. In return, errors are generated by the numerical integration of the energy function.
	
	The solution is sought in form of the displacement field $\boldsymbol{u}(\boldsymbol{X})$ that corresponds to the minimum total potential energy $\Pi$. This minimization can be accomplished by choosing the loss function $F$ to calculate the total potential energy $F := \Pi$.
	The input of the NN are points in the reference configuration $\boldsymbol X \in B \cup \Gamma$, in the domain $B$ or on its boundary $\Gamma$.
	
	A transformation is applied to integrate the geometric boundary conditions:
	Let the output of the NN be given by $\boldsymbol{z}(\theta, \boldsymbol{X})$. To retrieve the displacement field $\boldsymbol{u}_\theta(\boldsymbol{X})$ based on the parameter set $\theta$ of the NN, the displacements on the boundary are introduced in a separate term $\boldsymbol{u}_g(\boldsymbol{X})$. Additionally, a mapping $\boldsymbol{A}(\boldsymbol{X})$ with $\boldsymbol{A}(\boldsymbol{X})=\mathbf{0} \text { for } \boldsymbol{X} \in \Gamma$ is introduced. Then, the output is constructed as:
	\begin{align}
		\boldsymbol{u}_{\theta}(\boldsymbol{X})=\boldsymbol{u}_{g}(\boldsymbol{X})+\boldsymbol{A}(\boldsymbol{X}) \boldsymbol{z}(\theta, \boldsymbol{X})
		\label{e1-dem-1}
	\end{align}
	
	Now, $\boldsymbol{u}_\theta$ automatically fulfills the boundary constraints and a nonlinear optimization problem without constraints is obtained. This problem can be solved with an optimization procedure such as the Limited-memory Broyden-Fletcher-Goldfarb-Shanno (L-BFGS) algorithm. 
	
\begin{figure}
		\centering
		\includegraphics[width=\textwidth]{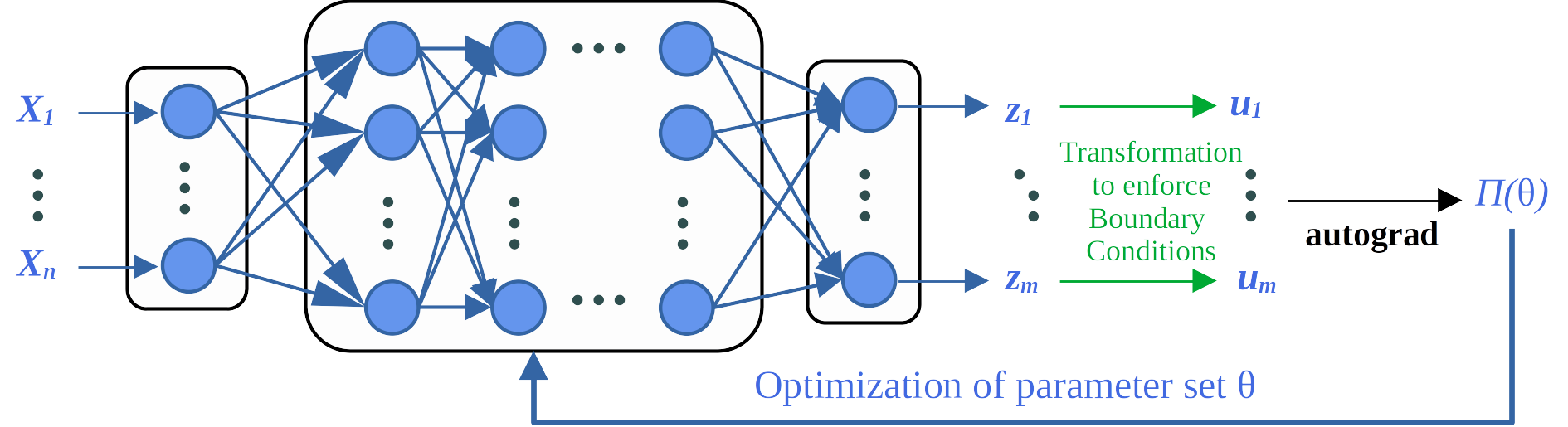}
		\caption[Information flow in DEM]{Information flow in DEM.}
		\label{infoflow-dem}
	\end{figure}

	The corresponding loss function to minimize is
	\begin{align}
		F^{\mathrm{DEM}}\left(\boldsymbol{u}_{\theta}\right)=\int_{B} \left( W\left(\boldsymbol{F}_{\theta}\right)-\boldsymbol{b} \cdot \boldsymbol{u}_{\theta} \right) \, \mathrm{d} V-\int_{\Gamma_{N}} \overline{\boldsymbol{T}} \cdot \boldsymbol{u}_{\theta} \, \mathrm{d} A
		\label{eq:f_dem}
	\end{align}
	with the elastic strain energy $W$, the deformation gradient $\boldsymbol{F}_\theta = \frac{\partial(\boldsymbol{X} + \boldsymbol{u}_\theta)}{\partial \boldsymbol{X}}$, body forces $\boldsymbol{b}$ and the traction $\overline{\boldsymbol{T}} = \boldsymbol{P}\cdot\boldsymbol{N}$ defined as the $1^{st}$ Piola-Kirchhoff stress tensor $\boldsymbol{P}$ projection on the outward normal of $\Gamma$. The deformation gradient $\boldsymbol{F}_{\theta}$ contains only first derivatives and is retrieved by the autograd feature of the NN framework. 
	The described procedure is shown in Fig. \ref{infoflow-dem}.

	Several numerical methods to calculate an approximation of the  integrals in the loss function have been suggested \cite{Nguyen-Thanh2020}. Some examples are the Monte Carlo integration and the trapezoidal rule, that we use for the examples presented in Section \ref{sec:Application-problems}.

	An alternative variant of the same concept is the Shallow Energy Method (SEM), where the deep NN is replaced by a shallow NN with a single hidden layer, activated by Radial Basis Functions (RBF) \cite{Liang2022}.
	
	An other enhancement of the basic DEM is the mixed DEM (mDEM) \cite{FuhgBouklas2022}, where both displacements $\boldsymbol{u}_{\theta}$ and stresses $\boldsymbol{P}_{\theta}$ are calculated by the NN. The deviation from the constitutive law  derived from the strain energy $W$ must then be integrated into the loss function to represent the correct material behavior 
	
	\begin{align}
		F^{\mathrm{mDEM}}=F^{\mathrm{DEM}}+\frac{V}{N_{f}} \sum_{j=1}^{N_{f}}\left|\left|\boldsymbol{P}_{\theta}\left(\boldsymbol{X}_{j}^{f}\right)-\left.\frac{\partial W\left(\boldsymbol{F}_{\theta}\right)}{\partial \boldsymbol{F}}\right|_{\boldsymbol{X}_{j}^{f}}\right|\right|_{2}^{2} \quad .
	\end{align}
	
	The prescribed forces on the Neumann boundary part $\Gamma_{N}$ can be accounted for directly by a transformation similar to Eq. \eqref{e1-dem-1} holding for the geometric boundary conditions on the Dirichlet boundary part $\Gamma_{D}$.
	Alternatively, an additional error term can be included to penalize the squared deviation from the prescribed forces.

	\subsection{competitive PINN (cPINN)}
	cPINNs extend the idea of PINNs by formulating the learning problem as a zero-sum game in the style of a Generative Adversarial Network (GAN) \cite{Zeng2022} with a Nash equilibrium that corresponds to the analytical solution of the PDE. This avoids the use of the square of the residual, which aims to improve the learning performance. 
	Compared to a classical PINN, cPINN introduces an additional discriminator FCNN with NN parameter set $\phi$ which is trained to predict errors of the PINN. 
	Let $\boldsymbol{u}_{\theta}$ be furthermore the output of the PINN and additionally $\boldsymbol{d}_{\phi}=\left(\boldsymbol{d}_{\phi}^{B}, \boldsymbol{d}_{\phi}^{\Gamma}\right)$ the output of the discriminator network. Then the minimax formulation of the game is given by
	
	\begin{align}
		\max _{\phi} \min _{\theta} F^{\mathrm{cPINN}}\left(\boldsymbol{u}_{\theta}, \boldsymbol{d}_{\phi}\right)=\left.\frac{1}{N_{f}} \sum_{i=1}^{N_{f}} \mathcal{N}\left[\boldsymbol{u}_{\theta}\right] \cdot \boldsymbol{d}_{\phi}^{B}\right|_{\boldsymbol{X}_{i}^{f}}+\left.\frac{\lambda_{b}}{N_{b}} \sum_{i=1}^{N_{b}} \mathcal{B}\left[\boldsymbol{u}_{\theta}\right] \cdot \boldsymbol{d}_{\phi}^{\Gamma}\right|_{\boldsymbol{X}_{i}^{b}} \,\,\,\, .
		\label{minimax-formulation}
	\end{align}
	
	\citeauthor{Zeng2022} \cite{Zeng2022} solve this optimization problem by using the Adam based Competitive Gradient Descent (ACGD) method. 

	\section{Neural Operator Methods} \label{subsec:neural-operator-methods}
	In a multi-query context, where a PDE must be evaluated for a large number of parameters, classical methods are computationally intensive. This includes both the conventional FEM and the Neural FEM methods explained previously. That drawback has motivated a large body of work on model order reduction, which deals with the tradeoff towards reduced accuracy, stability, and generalizability. Learning solution operators between (infinite-dimensional) function spaces using NNs is a comparatively young field.
	
	In this class of methods, the NN approximates the solution operator of
	the PPDE, i.e. for a given parameter set, the NN shall output the solution of the PDE at the points of interest. Now, the objective is defined as a risk functional that takes the probability distribution $\chi$ of the parameter set $\boldsymbol{y}$ into account \cite{Li2020b}
	\begin{align}
		F=\int_{\chi} \mathcal{L}\left(G_{\theta}(\boldsymbol{y}), G(\boldsymbol{y})\right) \mathrm{d} \chi \,\,,
	\end{align}
	where $G$ is the solution operator and $G_{\theta}$ its approximation by the NN. 
	
	A possibility to represent operators by means of NNs is a finite-dimensional approximation of the function spaces and interpolation of these spaces by the NN. Let e.g. the Boundary Value Problem (BVP) be given with approximations to the solutions calculated by traditional FEM on the node points.
	Now, the solution operator can be sought that maps the volume forces to the displacements on the node points, hence training a discrete operator \cite{Bhattacharya2020, Zhu2018}.
	However, this approach introduces a grid-dependency since the results are only obtained on the initially chosen node points. 
	
	Alternatively, the points of interest in space can be included as input parameter into the mapping that shall be represented by the NN,	
	\begin{align}
		\boldsymbol{u}_{\theta}: \quad B \cup \Gamma \times  \mathbb{R}^{\left(d+N_{\text {dof}}\right)} \rightarrow \mathbb{R}^{N_{\text {dof}}} \,\, , \quad[\boldsymbol{X}, \mathbf{y}] \mapsto \boldsymbol{u}_{\theta}(\boldsymbol{X} , \mathbf{y}) \,\,,
	\end{align}
	where $\mathbf{y}$ is taken as the parameter of the displacement field $\boldsymbol{u}_{\theta}$.

	In the following, Deep Operator Network (DeepONet) and Fourier Neural Operator (FNO) are discussed. Both approaches are discretization independent and allow for small generalization errors.

	\subsection{Deep Operator Network (DeepONet) and Physics Informed DeepONet (PIDeepONet)}
\begin{figure}
		\centering
		\includegraphics[width=0.9\textwidth]{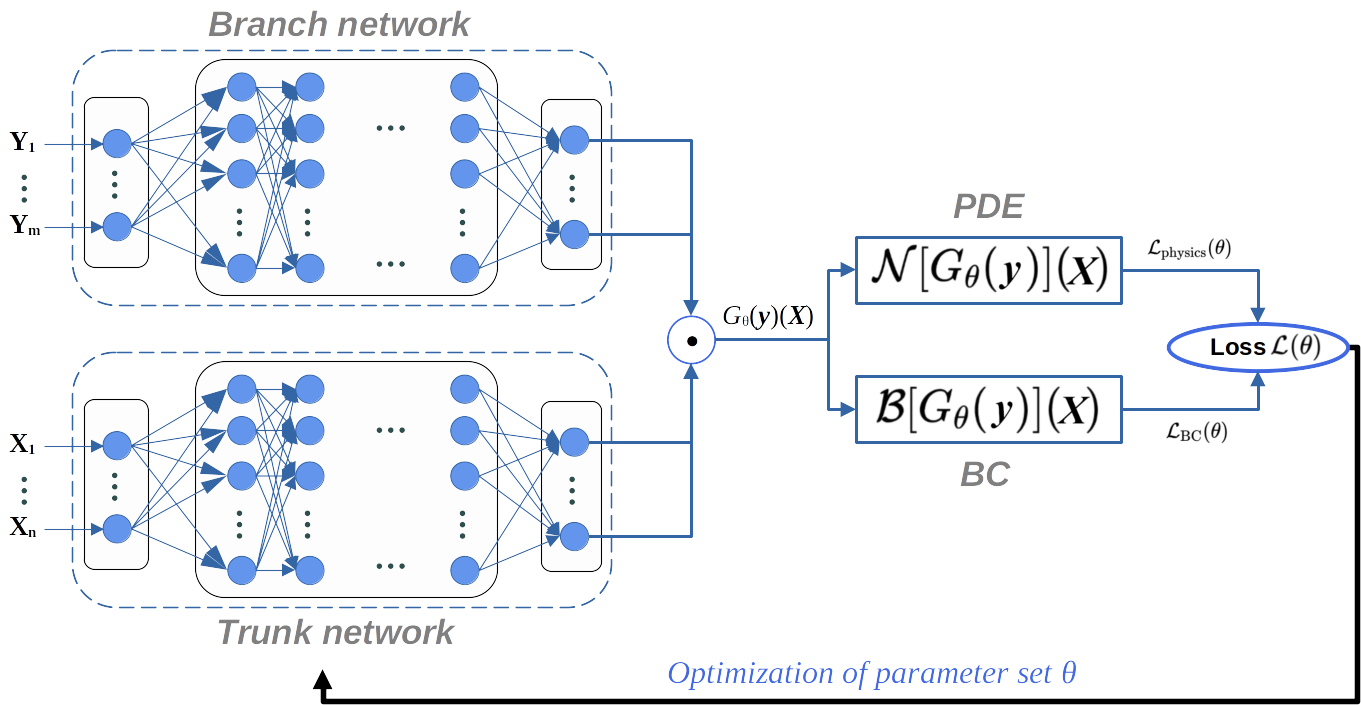}
		\caption{Information flow in PIDeepONet. Branch and trunk network are linked by the scalar product.}
		\label{PIDeepONet}
	\end{figure}

	NN can be employed as universal approximators of continuous functions, as well as for nonlinear continuous operators \cite{Wang2021b, Lu2021}
	
	\begin{align}
		\quad|G(\boldsymbol{y})(\boldsymbol{X})-\underbrace{\boldsymbol{g}\left(\boldsymbol{y}\left(\boldsymbol{X}_{1}\right), \ldots, \boldsymbol{y}\left(\boldsymbol{X}_{m}\right)\right)}_{\text {branch}} \cdot \underbrace{\boldsymbol{f}(\boldsymbol{X})}_{\text {trunk}}|<\varepsilon,
	\end{align}

	where $\boldsymbol{f}$ (trunk) and $\boldsymbol{g}$ (branch) can be represented by various classes of neural networks that satisfy the requirements of the classical universal approximation theorem \cite{Hornik1989}. It is assumed that the parameter $\boldsymbol{y}$ is known on sufficiently many grid points $m$. On this basis, the stacked and unstacked Deep Operator Network (DeepONet) are proposed in \cite{Lu2021}. The stacked DeepONet differs from the unstacked one only in the definition of the branch networks. In the unstacked DeepONet, these are combined into one net to facilitate training (Fig. \ref{PIDeepONet}).
	
	Let $\boldsymbol{y}=\left[\boldsymbol{y}\left(\boldsymbol{X}_{1}\right), \ldots, \boldsymbol{y}\left(\boldsymbol{X}_{m}\right)\right]^{\top}$ be the parameter function at discrete points in space. Moreover, let the output of DeepONet be $G_{\theta}(\mathbf{y})(\boldsymbol{X})$. Then, the empirical risk functional for DeepONet is given by
	
	\begin{align}
		F^{\text{DeepONet}}=\frac{1}{P \, N} \sum_{i=1}^{N} \sum_{j=1}^{P}\left(G_{\theta}\left(\boldsymbol{y}_{i}\right)\left(\boldsymbol{X}_{j}\right)-G\left(\boldsymbol{y}_{i}\right)\left(\boldsymbol{X}_{j}\right)\right)^{2} \quad .
	\end{align}

	Here, $N$ is the number of realizations of the parameter input $\boldsymbol{y}$ available for training and $P$ is the number of training data known per realization of the input function. The reference solution $G\left(\boldsymbol{y}_{i}\right)$ is determined by FEM-simulations or measurements.
	A single data point of the training data set consists of a triple of the form $(\boldsymbol{y}, \boldsymbol{X}, \boldsymbol{u}(\boldsymbol{X}))$. If $P>1$, the discrete parameter $\boldsymbol{y}$ must be repeated an appropriate number of times in the data set. For $N$ realizations of the parameter and $P$ evaluations per realization, the training dataset has a total of $N \times P$ entries.
	
	Despite of its simple structure, DeepONet can represent a wide range of mappings (i.e. it is very expressive) and allows to achieve small generalization errors. Furthermore, it can be applied very easily to arbitrary parametrizations.
	
	The physics informed variant PIDeepONet \cite{Wang2021b} extends the empirical risk functional by adding a physically motivated term ensuring the compliance with the PDE in the weak form. For this purpose, the risk functional is extended by adding the squared residuals of the (nonlinear) differential operators $\mathcal{N}, \mathcal{B}$

	\begin{equation}
		\begin{aligned}
			F^{\text {PIDeepONet }}=F^{\text {DeepONet }} & +\frac{1}{m N_{y}} \sum_{i=1}^{N_{y}} \sum_{j=1}^{m} \mathcal{N}^{2}\left[G_{\theta}\left(\mathbf{y}_{i}\right)\right]\left(\boldsymbol{X}_{j}\right) \\
			& +\frac{1}{N_{b} N_{y}} \sum_{i=1}^{N_{y}} \sum_{j=1}^{N_{b}} \mathcal{B}^{2}\left[G_{\theta}\left(\mathbf{y}_{i}\right)\right]\left(\boldsymbol{X}_{j}\right) .
		\end{aligned}
		\label{DeepONet-extended-Res}
	\end{equation}
	In return, no FEM reference data is necessary.
	
	Eq. \eqref{DeepONet-extended-Res} is formulated for the case, where the residual can only be evaluated at the $m$ discretization points, where the parameters $\boldsymbol{y}$ are known.  
	Alternatively, an extended data set can be generated which contains $\left\{\boldsymbol{y}(X_{1}), \ldots, \boldsymbol{y}(X_{m}) , X_{1}, \boldsymbol{y}(X_{1}), \ldots,  X_{r}, \boldsymbol{y}(X_{r})\right\}$ with $r$ the number of (randomly determined) additional gridpoints. 
	
	\subsection{Fourier-Neural-Operator (FNO)} \label{subsec:FNO}
	An FNO \cite{Li2020b} represents the solution operator of a
	PPDE with the help of a series of Fourier Blocks. 
	
\begin{figure}
		\centering
		\includegraphics[width=0.75\textwidth]{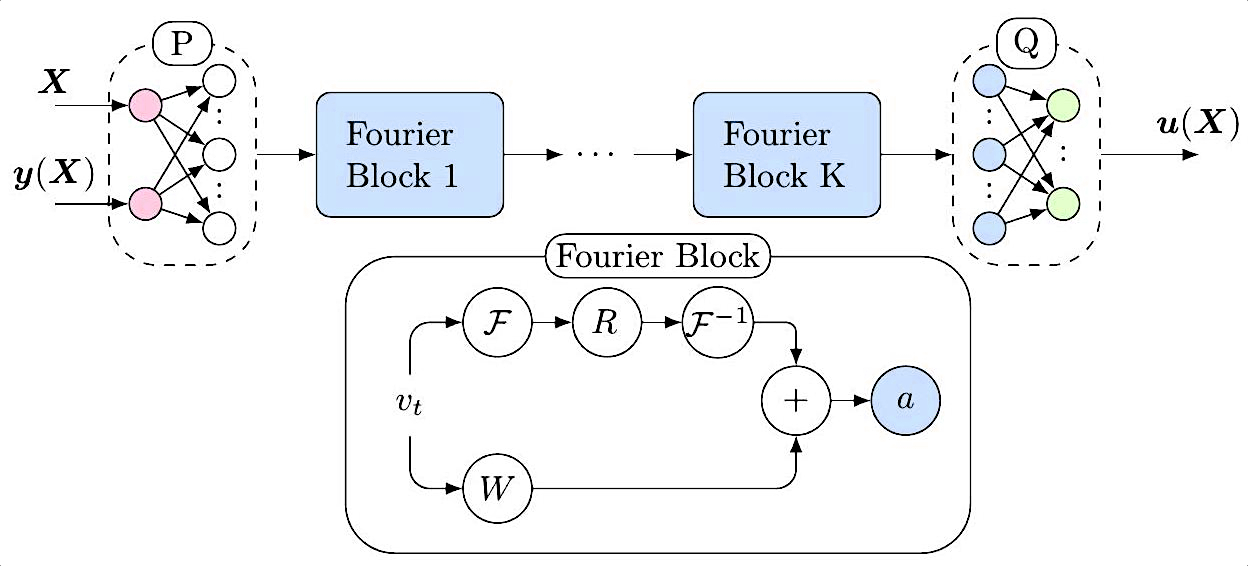}
		\caption[FNO schema]{Schematic representation of the FNO (adapted from \cite{Li2020b}). Layers with nonlinear activation (GELU) are marked in blue.
		}
		\label{fno-schema}
	\end{figure}

	A Fourier block includes several operations  as shown in Fig. \ref{fno-schema}: i) It applies the Fourier transform $\mathcal{F}$ (in form of the Fast Fourier Transform, FFT) on its input $\boldsymbol{v}^t (\boldsymbol{X})$. ii) It applies a linear transform $R_\theta^t$ (parameterized by the NN parameter set $\theta$) on the lower Fourier modes and filters out the higher modes. iii) It applies the inverse Fourier transform $\mathcal{F}^{-1}$. iv) In parallel, the Fourier block applies another linear transform $W_\theta^t$ on the input $\boldsymbol{v}^t (\boldsymbol{X})$. v) Results of both branches are summed up and forwarded to the nonlinear activation function $\sigma$. 
	
	The input $[\boldsymbol{X}, \boldsymbol{y}(\boldsymbol{X})] \in \mathbb{R}^{N_{dof} + d}$ of the network is lifted up to a higher dimension $d_v$ by a shallow (e.g. single-layer) FCNN $P$ with linear activation function. Another FCNN $Q$ projects the output of the last Fourier Block onto the output space $\mathbb{R}^{N_{dof}}$ which results in the FNO output $\boldsymbol{u}_\theta(\boldsymbol{X})$. The whole process is visualized in Fig. \ref{fno-schema} and can be described by the iterative architecture
	
	\begin{equation}
		\begin{aligned}
			\boldsymbol{v}_{0}(\boldsymbol{X}) & =P(\boldsymbol{y}(\boldsymbol{X}), \boldsymbol{X}) \\
			\boldsymbol{v}_{t+1}(\boldsymbol{X}) &=\sigma\left(W_{\theta}^{t} \boldsymbol{v}_{t}(\boldsymbol{X})+\mathcal{F}^{-1}\left[R_{\theta}^{t} \cdot \mathcal{F}\left[\boldsymbol{v}_{t}\right]\right](\boldsymbol{X})\right)\\
			\boldsymbol{u}_{\theta}(\boldsymbol{X}) & =Q\left(\boldsymbol{v}_{K}(\boldsymbol{X})\right) \,,
		\end{aligned}
	\end{equation}
	where $K$ is the number of sequential Fourier Blocks.

	\subsubsection{Physically Informed Neural Operator (PINO)}
	An extension of the FNO to a physically informed neural operator (PINO) \cite{Li2022} has been investigated as well. FNO computes displacements on an equidistant grid, so that a DEM-like extension is straight forward. Therefor, the same potential energy as for the DEM (see Eq. \ref{eq:f_dem}) was added to the loss function
	\begin{align}
		F^{\text {PINO}}=F^{\mathrm{FNO}}+F^{\mathrm{DEM}} = F^{\mathrm{FNO}}+  \int_{B} \left( W\left(\boldsymbol{F}_{\theta}\right)-\boldsymbol{b} \cdot \boldsymbol{u} \right) \, \mathrm{d} V-\int_{\Gamma_{N}} \overline{\boldsymbol{T}} \cdot \boldsymbol{u} \, \mathrm{d} A
		\label{eq:4-35} \quad.
	\end{align}

	\section{Application to examples in elastostatics} \label{sec:Application-problems}
	The present section compares the NN methods previously explained by means of example boundary value problems, including several 1D examples and one 2D example with two load cases.
	
	\subsection{The 1D tensile bar}
\begin{figure}
		\centering
		\includegraphics[width=0.65\textwidth]{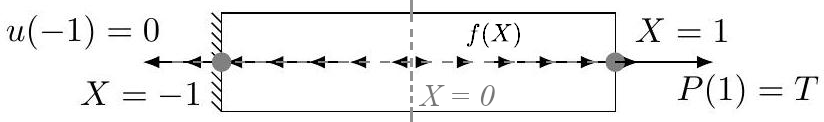}
		\caption{Illustration of the 1D BVP with BC: $u(X=-1)=0$ and $P(X=1)=T$ .}
		\label{fig:4-1}
	\end{figure}

	The first example is based on the setup shown in Fig. \ref{fig:4-1}. The bar is
	clamped at the left edge $u(-1)=0$ and loaded along its entire length with the force density $f(X)$. A Neumann boundary condition $P(1)=T$ is applied at the right edge. 
	
	\subsubsection{Example A} \label{subsubsec:ProblemA}
	The following energy density is considered
	\begin{align}
		W(F)=F^{\frac{3}{2}}-\frac{3}{2} F+\frac{1}{2} \quad \text { with } \quad F=1+u^{\prime}(X) \,\, .
		\label{eq:4-1}
	\end{align}
	
	From Eq. \eqref{eq:4-1}, the first Piola-Kirchhoff stress reads
	\begin{align}
		P=\frac{\partial W}{\partial F}=\frac{3}{2}\left(F^{\frac{1}{2}}-1\right) \Rightarrow-\frac{\partial P}{\partial X}=-\frac{3}{4} \frac{1}{\sqrt{1+u^{\prime}}} u^{\prime \prime}(X)=f(X) \,\,.
		\label{eq:4-2}
	\end{align}
	
	Specifically, the force density $f(X)=X$ is chosen and the load at the free end is set to zero: $T=0$
	\begin{align}
		-\frac{3}{4} \frac{1}{\sqrt{1+u^{\prime}}} u^{\prime \prime}(X)=X \quad \text { with } \quad u(-1)=0, \,\, T=0 \Rightarrow u^{\prime}(1)=0 \, .
		\label{eq:4-14}
	\end{align}
	
	The example has the following analytical solution which will be used to validate the results obtained by the Neural FEM methods 
	\begin{align}
		u(X) & =\frac{1}{135}\left(3 X^{5}-40 X^{3}+105 X+68\right)
		\label{eq:4-15}
		\\
		u^{\prime}(X) & =\frac{1}{9}\left(X^{4}-8 X^{2}+7\right) \,\, .
		\label{eq:4-16}
	\end{align}

	\subsubsection{Example B} \label{subsubsec:ProblemB}
	A linear elastic material is investigated:
	\begin{align}
		W=\frac{1}{2}\left(u^{\prime}\right)^{2} \quad \Rightarrow \quad P(X)=u^{\prime}(X) \quad \Rightarrow \quad-u^{\prime\prime}(X)=f(X)
		\label{eq:4-3}
	\end{align}
	
	\paragraph{Example B1}
	A single load case is analyzed in examples related to PINN and DEM.
	\begin{align}
		-E u^{\prime \prime}(X)= f(X) = Q \cdot A \quad \text { with } \quad u(-1)=0 \,\, \text{ and } \,\, E u^{\prime}(1)=T.
		\label{eq:4-4}
	\end{align}
	Here, distributed forces take the values $Q = 9.395 \cdot 10^{4} \mathrm{~Nm}^{-1} \text{ and } T= 1.015 \cdot 10^{8} \mathrm{~Nm}$.
	Young's modulus corresponds to steel ($E = 210 \cdot 10^{9} \mathrm{~N} \mathrm{~m}^{-2}$) and the cross section surface is $A = 1 \mathrm{~m}^2$.

	\paragraph{Example B2} \label{par:Problem-B2}
	For the Neural Operator models, the PPDE is normalized and a parameterization of the force density $f$ as well as a parameterization of the Neumann boundary condition are studied. Since the boundary only consists of one point, the boundary condition can be described by a scalar $\pi_2$ for which a uniform distribution between $[0,1]$ is assumed. The reference solutions are computed using FEniCS. The BVP is described by

	\begin{align}
		-\frac{\partial^{2} u}{\partial X^{2}}=f(X) \quad \text { with } \quad\left\{\begin{array}{c}
			u(-1)=0 \\
			u^{\prime}(1)=\pi_{2}
		\end{array}\right. \,\,.
		\label{eq:4-33}
	\end{align}

	\subsection{The plate -- Example C} \label{subsubsec:4-2-4}
	The selected two-dimensional example deals with a plate made of a Neo-Hookean material with the energy density
	\begin{align}
		W(\boldsymbol{F})=\frac{\mu}{2}\left(I_{1}-2-\ln \mathsfit{J}\right)+\frac{\lambda}{2}(\mathsfit{J}-1)^{2} \,\,.
		\label{eq:4-23}
	\end{align}

	$I_{1}=\operatorname{tr}\left(\boldsymbol{C}\right)$ is the first invariant of the right Cauchy-Green deformation tensor $\boldsymbol{C} = \boldsymbol{F}^T \boldsymbol{F}$ and $\mathsfit{J}=\operatorname{det}(\boldsymbol{F})$ the determinant of the deformation gradient. The corresponding derivatives are
	$\frac{\partial \mathsfit{J}}{\partial \boldsymbol{F}}=\mathsfit{J} \boldsymbol{F}^{-1}$  and $\frac{\partial \operatorname{tr}\left(\boldsymbol{F}^{\top} \boldsymbol{F}\right)}{\partial \boldsymbol{F}}=2 \boldsymbol{F} $.
	The symbols $\lambda$ and $\mu$ denote the Lamé constants. 
	
	For the energy density, Eq. \eqref{eq:4-23}, the 1st and 2nd Piola-Kirchhoff stress tensor are given by:	
	\begin{align}
		\boldsymbol{P} =\frac{\partial W}{\partial \boldsymbol{F}}=\mu \boldsymbol{F}+(\lambda \ln \mathsfit{J}-\mu) \boldsymbol{F}^{-\top} \quad \text { and } \quad \label{eq:4-25} 
		\boldsymbol{S} =\boldsymbol{F}^{-1} \cdot \boldsymbol{P}=\mu \boldsymbol{I}+(\lambda \ln \mathsfit{J}-\mu) \boldsymbol{C}^{-1} \quad .
	\end{align}

	\begin{figure}
		\centering
		\includegraphics[width=0.5\textwidth]{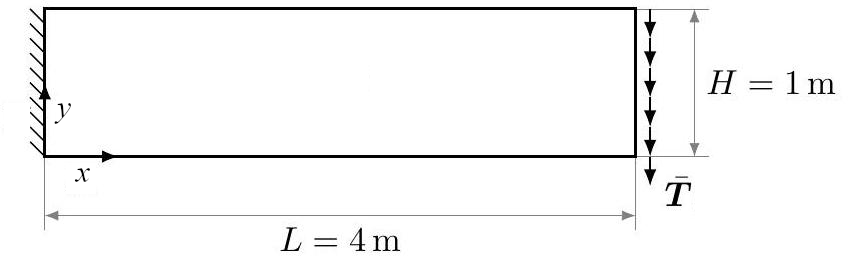}
		\caption{Geometry of the 2D BVP.}
		\label{fig:4-14}
	\end{figure}

	The studied example is shown in Fig. \ref{fig:4-14}. The plate is clamped at the left edge and the Neumann boundary conditions are set at the right edge. The components of the kinematic fields in Cartesian coordinates are given by
	
	\begin{align}
		[\boldsymbol{u}]=\left[\begin{array}{l}
			u_{x} \\
			u_{y}
		\end{array}\right] \quad \Rightarrow \quad [\boldsymbol{F}]=\left[\begin{array}{ll}
			F_{x x} & F_{x y} \\
			F_{y x} & F_{y y}
		\end{array}\right]=\left[\begin{array}{cc}
			1+u_{x, x} & u_{x, y} \\
			u_{y, x} & 1+u_{y, y}
		\end{array}\right] \quad .
		\label{eq:4-27}
	\end{align}
	
	From Eq. \eqref{eq:4-25}, the first and second Piola-Kirchhoff stress tensor are calculated as follows 
	\begin{align}
		[\boldsymbol{P}] &=\left[\begin{array}{cc}
			P_{x x} & P_{x y} \\
			P_{y x} & P_{y y}
		\end{array}\right]=\mu\left[\begin{array}{cc}
			F_{x x} & F_{x y} \\
			F_{y x} & F_{y y}
		\end{array}\right]+\frac{\lambda \ln \mathsfit{J}-\mu}{\mathsfit{J}}\left[\begin{array}{cc}
			F_{y y} & -F_{y x} \\
			-F_{x y} & F_{x x}
		\end{array}\right] \quad ,\\´
		[\boldsymbol{S}] &=\left[\begin{array}{ll}
			S_{x x} & S_{x y} \\
			S_{y x} & S_{y y}
		\end{array}\right]=\left[\begin{array}{cc}
			\mu & 0 \\
			0 & \mu
		\end{array}\right]+\frac{\lambda \ln \mathsfit{J}-\mu}{\mathsfit{J}^{2}}\left[\begin{array}{cc}
			C_{y y} & -C_{x y} \\
			-C_{y x} & C_{x x}
		\end{array}\right] \quad .
		\label{eq:4-29}
	\end{align}
	
	Moreover, an equivalent stress is calculated as in \cite{Nguyen-Thanh2020} and used in contour plots (Section \ref{subsubsec:cPINN-results})\\ $S_{E}=\sqrt{0.5 \left(\left(S_{x x}-S_{y y}\right)^{2}+S_{x x}^{2} + S_{y y}^{2}\right) +3 S_{x y}} \label{eq:4-30}$.\\ 
	Two load cases are investigated for the setup described. 
	
	\paragraph{Example C1}
	The first load case deals with the vertical load $\overline{\boldsymbol{T}}=-5 \boldsymbol{e}_{y}$ .
	
	\paragraph{Example C2}
	The second load case is uniaxial tension with $\overline{\boldsymbol{T}}=50 \boldsymbol{e}_{x}$.

	\subsection{Error measure}
	With the solution operator of the PPDE $G: \mathcal{Y} \rightarrow \mathcal{S}$ and its NN approximation $G_{\theta}$, the average relative $L_{2}$ error for the $N$ test data sets is calculated for the Neural Operator methods \cite{Bhattacharya2020, Li2020b, Lu2021} as:
	
	\begin{align}
		\epsilon_{\mathrm{rel}} = \frac{1}{N} \sum_{j=1}^{N} \frac{||G_{\theta}(\boldsymbol{y}_{j})-G(\boldsymbol{y}_{j})||_{L_{2}}}{||G(\boldsymbol{y}_{j})||_{L_{2}}} \quad .
		\label{eq:4-8}
	\end{align}
	
	With the Neural FEM, only one concrete BVP can be analyzed at once. Then, the relative $L_{2}$ error is calculated based on the solution for the displacement field
	 
	\begin{align}
		\epsilon_{\mathrm{rel}}=\frac{||\boldsymbol{u}_{\theta}-\boldsymbol{u}||_{L_{2}}}{||\boldsymbol{u}||_{L_{2}}} \quad .
		\label{eq:4-9}
	\end{align}

	In both cases, the determination of the relative $L_{2}$-error requires the computation of the $L_{2}$-norm which is approximated by the discrete $L_{2}$-norm. On an equidistant lattice $\left\{\boldsymbol{X}_{i}^{\text {equi }}\right\}_{i=1}^{N}$, the discrete $L_{2}$-norm is calculated as
	\begin{align}
		||\boldsymbol{f}||_{L_{2, d}}^{2}=\Delta V \sum_{i=1}^{N}||\boldsymbol{f}(\boldsymbol{X}_{i}^{\mathrm{equi}})||_{2}^{2}=\Delta V \sum_{i=1}^{N} \sum_{j=1}^{d} f_{j}^{2}(\boldsymbol{X}_{i}^{\mathrm{equi}}) \,\,,
		\label{eq:4-11}
	\end{align}
	with the volume (in 2D: surface area) of each lattice unit $\Delta V$.

	\subsection{Numerical integration}
\begin{figure}
		\begin{minipage}{0.49\textwidth}
			\centering
			\includegraphics[width=0.85\textwidth]{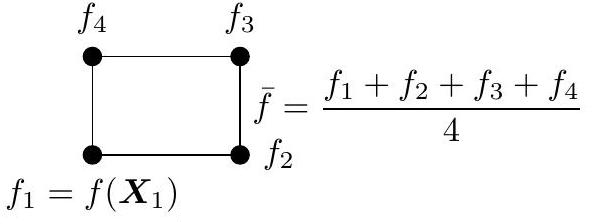}
		\end{minipage}
		\begin{minipage}{0.49\textwidth}
			\centering
			\includegraphics[width=0.75\textwidth]{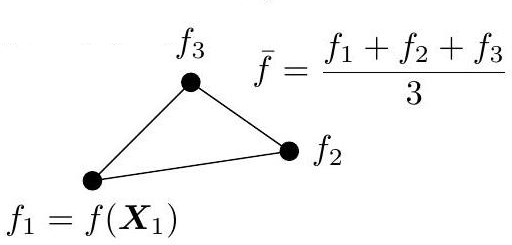}
		\end{minipage}
		\caption{Examples for polytopes.}
		\label{fig:polytopes}
	\end{figure}

	The discrete $L_{2}$-norm is based on a simple Riemann sum with error order $\mathcal{O}(\Delta V)$. For the approximation of the risk functional, e.g., in connection with the calculation of the potential energy in the DEM, other integration methods have to be considered. Two classical methods are the Monte Carlo (MC) integration and the trapezoidal rule. The trapezoidal rule requires partitioning of the integration domain into polytopes. In the simplest case, these would be hypercubes on an equidistant grid. 
	In \cite{FuhgBouklas2022}, an integration method based on the Delaunay triangulation is proposed. The trapezoidal rule still applies, where $\bar{f}_{i}$ is the average value over the $i$-th simplex (e.g. triangles). The two polytopes for integration in 2D are shown as examples in Fig. \ref{fig:polytopes}.	
	Let $V$ be the volume of the integration domain and $\bar{f}_{i}$ the average of $f$ over the corners of the $i$-th polytope with $i \in [1, N]$ and $N$ the number of vertices. Then, the integral approximations are generally given by Eq. (\ref{eq:4-12}a) for Monte-Carlo and Eq. (\ref{eq:4-13}b) for the trapezoidal rule. 
		
		\begin{align}
			I_{\mathrm{MC}}(f) =\frac{V}{N} \sum_{i=1}^{N} f\left(\boldsymbol{X}_{i}\right)
			\label{eq:4-12} \quad \text{(a)} &&\qquad
			I_{\mathrm{T}}(f)=\sum_{i=1}^{N} \bar{f}_{i} \,\, \Delta V_{i}
			\quad \text{(b)} \quad .
		\end{align}

\begin{figure}
		\centering
		\includegraphics[width=0.85\textwidth]{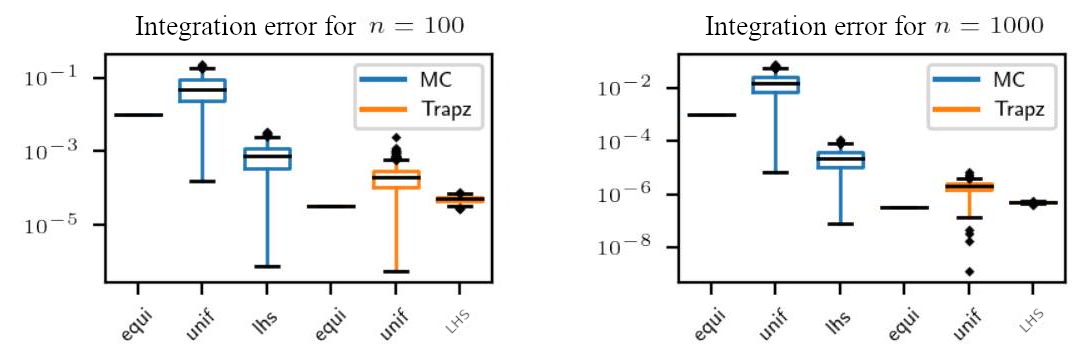}
		\caption{Comparison of Monte Carlo (MC) integration and the trapezoidal rule with $n=100$ and $n=1000$ grid points for equidistant sampling (equi), pseudo-random uniform sampling (unif) and Latin Hypercube Sampling (LHS).}
		\label{fig:4-3}
	\end{figure}

\begin{figure}
		\centering
		\includegraphics[width=0.6\textwidth]{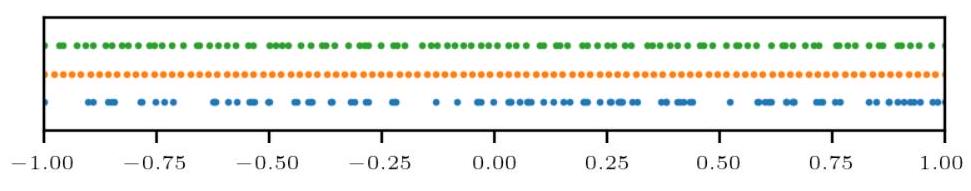}
		\caption{Quasi-random (top), pseudo-random (bottom), and equidistant (middle) grid points for numerical integration (1D case).}
		\label{fig:4-2}
	\end{figure}

	Three methods are investigated to select the grid points: equidistant grid points, pseudo-random numbers and quasi-random numbers (Latin Hypercube Sampling, LHS). Exemplarily, we compare the absolute error of the potential energies in the nonlinear 1D setup (Example A). The trapezoidal rule with 100 000 grid points is assumed as a quasi-exact comparison value. The MC integration with equidistant grid points reduces to a simple Riemann sum.  The results for 100 and 1000 grid points, respectively, are summarized in Fig. \ref{fig:4-3}. A characteristic distribution of grid points is shown in Fig. \ref{fig:4-2}.
	
	Due to the larger integration error, uniform pseudo-random sampling is not considered further. In the following, "random" sampling always refers to quasi-random LHS. Fig. \ref{fig:4-3} also shows that the trapezoidal rule is consistently more accurate than MC integration. 
	
	\subsubsection{Technical Implementation}
	In this work, we use PyTorch (version: 1.11.0), which contains the optimizer L-BFGS that is employed in some of the investigated methods. The computations have been run on an Intel Core i5-7200U mobile processor. A mobile NVIDIA GeForce GTX 950m is used as graphics card.
	
	The parameters of the NN are always randomly initialized, so that the results underly statistical variations.
	
	\subsection{Neural FEM}
	\subsubsection{PINN}
	
\begin{figure}
		\centering
		\includegraphics[width=0.8\textwidth]{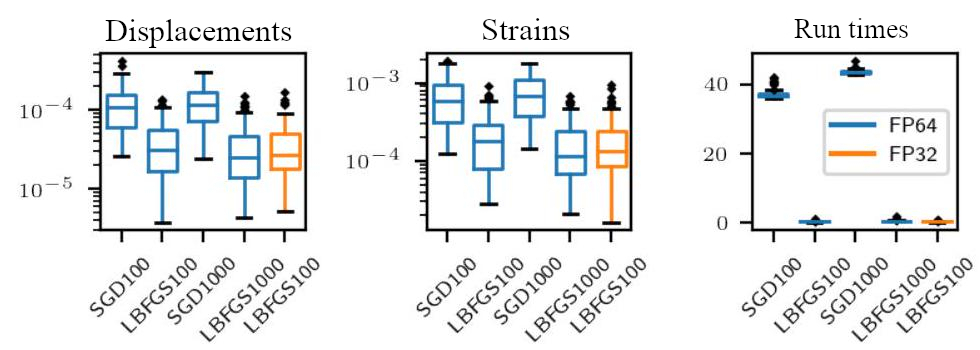}
		\caption{Relative $L_{2}$ error and run times in s for two optimization methods L-BFGS, SGD with 100 and 1000 collocation points and FP32/ FP64 accuracy.}
		\label{fig:4-5}
	\end{figure}

	\paragraph{Example A}
	In the numerical experiments related to Example A, the NN architecture $[1,10,1]$ is always used. Two optimizers (L-BFGS, SGD), different numbers of collocation points (100 and 1000 points) and different computational accuracies (single precision FP32 and double precision FP64) are compared (Fig. \ref{fig:4-5}).
	
	The SGD is repeated for 10000 epochs and the L-BFGS for 15 epochs, each with the default parameters of the methods. 
	The discrepancy in the required number of epochs is reflected by the run time, which is about $43 \mathrm{~s}$ for the SGD compared to about $0.300 \mathrm{~s}$ for the L-BFGS. Obviously, a second-order method (such as the L-BFGS) can greatly reduce the number of necessary iterations. In each epoch, the complete data set is used (Full-Batch). Despite the same information being provided to the NN, the L-BFGS method performs better on average than the SGD. Fig. \ref{fig:4-5} also shows that the reduction in total error relative to the number of collocation points quickly goes into saturation. The difference between $N=100$ and $N=1000$ is only about $6.500 \cdot 10^{-6}$ for the L-BFGS. 
	
	Moreover, no significant increase of the total error measured in the relative $L_{2}$ norm is shown when computing on single precision. This can greatly reduce the computation time on commercially available graphics cards that are optimized for single precision computing. However, this needs to be confirmed in the further research for more complex problems.
	
\begin{figure}
		\centering
		\begin{subfigure}{0.49\textwidth}
			\includegraphics[width=0.8\textwidth]{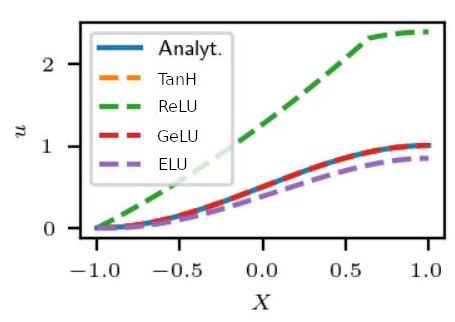}
			\caption{Displacements}
		\end{subfigure}
		\begin{subfigure}{0.49\textwidth}
			\includegraphics[width=0.8\textwidth]{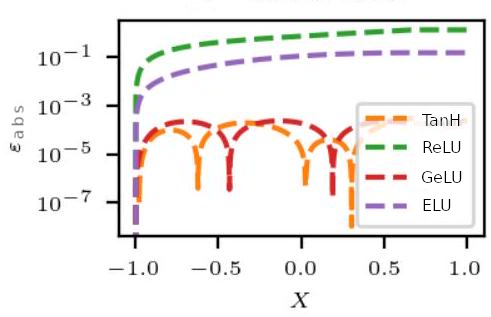}
			\caption{Absolute error}
		\end{subfigure}
		\caption{Calculated displacements with 100 collocation points and different activation functions.}	\label{fig:4-6}
	\end{figure}

	The best results were obtained with the tangent hyperbolic (Tanh) activation function. Other activation functions, such as the Rectified Linear Unit (ReLU) or the Exponential Linear Unit (ELU) do not converge or converge very poorly against the analytical solution of the problem. The calculated displacements for different activation functions are comparatively shown in Fig. \ref{fig:4-6}. ReLU and ELU could not be optimized with L-BFGS. Therefore, only the results after optimization with Adam are shown. 
	The second derivative of the approximation with ReLU activation is everywhere constantly zero, (except for the point at the kink). This destroys the information in the residual and a training of the network must necessarily fail. Hence, for the following studies, Tanh was always applied as the activation function. 
	Other activation functions are not considered in the present contribution. However, in the literature, the composite function $\max \left(0, x^{3}\right)$ \cite{EandYu2018}, the Swish activation $z S(\beta z)$ (where $S$ denotes the sigmoid activation function) \cite{Karumuri2020} and GELU \cite{Li2020b} have been successfully employed. 
	
	The training of PINN with SGD took about $40 \mathrm{~s}$ for 10 000 epochs.

	\paragraph{Example C}\label{par:DCM-ProblemC}
	Amongst the conventional PINN representatives, the DCM is the easiest to implement for 2D problems and thus chosen to apply to Example C (Section \ref{subsubsec:DCM}).
	
	Within the domain, the balance of linear momentum reads $\nabla \cdot \boldsymbol{P}=\mathbf{0}$, which already is a residual form for approximations of $\boldsymbol{P}$. The Neumann boundary conditions are given by $\boldsymbol{P} \cdot \boldsymbol{N}=\overline{\boldsymbol{T}}$, the Dirichlet constraints are incorporated directly by the application of a transformation on the output of the NN (Eq. \eqref{e1-dem-1}). 
	
	The architecture of the network is specified as $[2,30,30,2]$ and on the Neumann boundary part, 900 random collocations points are chosen. 
	4000 collocation points are used within the body. L-BFGS with learning rate 1.0 and Line Search with Wolfe condition is applied as optimizer. 
	
	Other than the DEM, the DCM does not converge towards the reference solution for load case C1.
	A comparison with the DEM shows that the boundary conditions are not appropriately learned by the DCM.

\begin{figure}
		\centering
		\includegraphics[width=\textwidth]{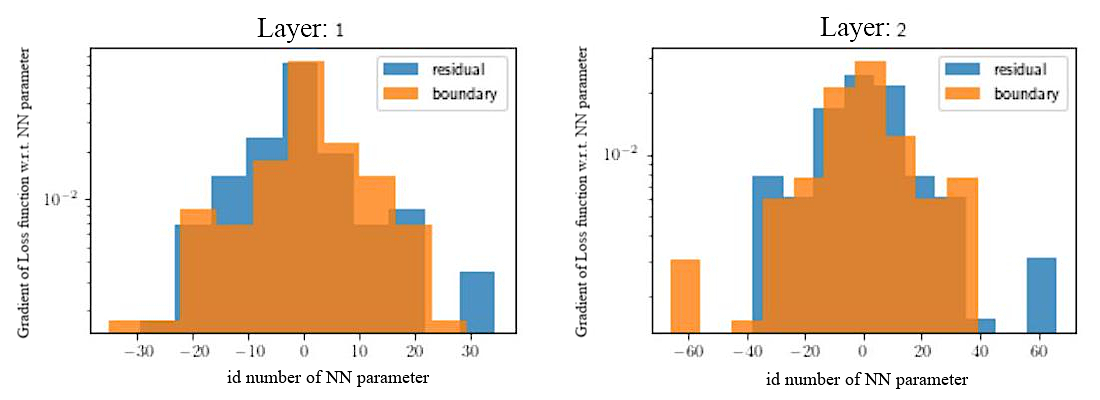}
		\caption{Histogram over the gradients of empirical risk. First hidden layer (left), output layer (right).}
		\label{fig:4-18}
	\end{figure}

	\citeauthor*{Wang2021a} \cite{Wang2021a} discuss that the training of PINNs may fail due to numerical inaccuracies if the contributions to the loss value -- one portion from the residual and the other portion from the Neumann boundary part -- or their gradients w.r.t to the NN parameters are in vastly different orders of magnitude.
	In the case of Example C, the loss value in the DCM consists of the portion from the residual with the value $0.119$ and the portion from the Neumann boundary part with the value $1.121$.
	The gradients of each contribution w.r.t. the NN parameters are relatively uniformly distributed (Fig. \ref{fig:4-18}).
	Moreover, the correct boundary conditions are not learned even if the loss portion of the residual is excluded (manually set to 0). 
	Hence, the failure of the DCM on the example cannot be explained by this kind of numerical inaccuracies. 	
	It is possible that the optimizer gets stuck in a local minimum, but this behavior needs a further investigation.
	
	\subsubsection{cPINN} \label{subsubsec:cPINN-results}
	\paragraph{Example A}
	For many applications, the accuracy that can be achieved with a classical PINN is not sufficient. cPINN was developed to improve the accuracy by avoiding the squaring of the residual \cite{Zeng2022, Schaefer2019}. Instead, it employs Adaptive Competitive Gradient Descent (ACGD) as optimization procedure whose Python implementation is publicly available \cite{cgds-package}. Furthermore, Tanh in the PINN part and ReLU in the discriminator network are chosen as the activation functions.
	
	The architecture of the PINN is chosen with 10 neurons in the hidden layer (architecture: $[1,10,1]$ ) for comparability with the conventional PINN/ DCM. The layer width $h$ of the discriminator, on the other hand, was varied (architecture: $\left[1, 20, 2\right]$ vs. $\left[1, 50, 2\right]$).
	Initial experiments have shown that the output of the discriminator $\boldsymbol{d}_{\phi}$ must be separated for points in the domain and on the boundary $\boldsymbol{d}_{\phi}=\left(\boldsymbol{d}_{\phi}^{B}, \boldsymbol{d}_{\phi}^{\Gamma}\right)$.
	Therefore, the output layer contains 2 neurons. The option of separating both outputs of the discriminator into independent subnetworks is also tested. However, this did not result in any improvement. Based on these results, only the first variant with the smaller number of NN parameters is considered further. All calculations are performed with double precision.
	
	\begin{figure}
		\centering
		\includegraphics[width=0.8\textwidth]{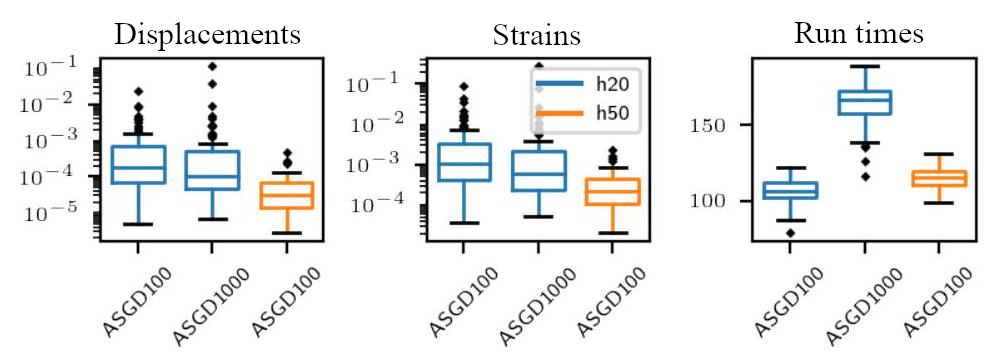}
		\caption{Relative $L_{2}$ errors and run times in s of cPINN with 100 and 1000 collocation points and 20 and 50 neurons in the hidden layer of the discriminator.}
		\label{fig:4-7}
	\end{figure}
	
	Fig. \ref{fig:4-7} shows the results for accuracy and run times, from 100 runs with random NN parameter initializations in form of a box plot. It can be seen that the accuracy is only moderately affected by the number of collocation points. However, the width of the discriminator has a significant impact on the training result. Increasing the number of neurons in the hidden layer from 20 to 50 reduces the error by an order of magnitude. It is not entirely clear why such a large discriminator network is necessary. Moreover, the training is relatively slow, taking about 2 min (up to 3 min for 1000 collocation points) for about 6000 epochs. 
	
 	The improvement of up to 2 orders of magnitude reported by \cite{Zeng2022} cannot be demonstrated here. This may be because the pathologies related to training PINNs \cite{Krishnapriyan2021, Wang2021a}, which cPINN addresses, do not arise in this simple example. On the other hand, the regularization by the residual leads to a complex energy landscape of the optimization procedure, making optimization more difficult \cite{Krishnapriyan2021}. 
	Moreover, the material law sometimes causes the optimization process to abort if the network parameters have been initialized unfavorably. This problem can be solved by reducing the range for random sampling of the initial parameter values.
	
\begin{figure}
		\centering
		\includegraphics[width=\textwidth]{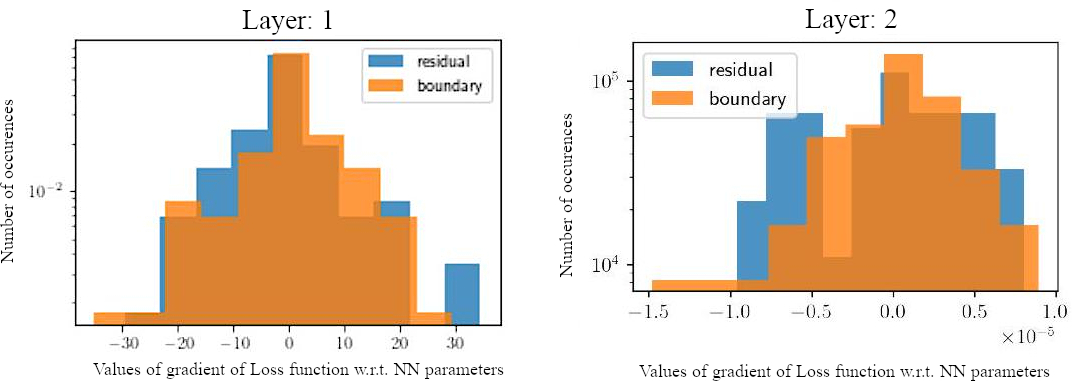}
		\caption{Histogram of the gradients of the optimizable parameters. First hidden layer (left), output layer (right).}
		\label{fig:4-9}
	\end{figure}

	In addition, the different weighting of the summands of the empirical risk 
	can lead to different magnitudes of the gradients of the loss function w.r.t. the NN parameters, which can impair the NN parameter optimization \cite{Wang2021a, Wang2021b}. In the present case, the gradients for the residual within the domain $\nabla_{\theta} F^{\mathrm{PINN}, B}$ are much larger than the gradients for the constraints $\nabla_{\theta} F^{\mathrm{PINN}, \Gamma}$.
	The optimization procedure is therefore driven more strongly toward a solution that reduces the residual while allowing for deviations from the constraints. As a result, the optimization procedure converges toward a plausible solution, but one that does not satisfy the boundary conditions. For the complex architecture $[1,50,50,1]$, the gradients of the residual and boundary portion of the loss function w.r.t. the NN parameters of the first hidden layer and the output layer are shown in Fig. \ref{fig:4-9}. However, not much discrepancy is detected in the order of magnitude of the gradient values for both portions of the loss function.

	\subsubsection{DEM}
	With the architecture $[1,10,1]$, DEM training runs on average twice as fast compared to classical PINN. For deeper networks, this fact is amplified, as will be shown in the analysis of Example C. 
	Our own DEM implementation is based on the public source code \cite{Nguyen-Thanh2020} and is extended to allow training with Monte Carlo integration on quasi-random grid points. 
	Tanh is used as the activation function. 
	
\begin{figure}
		\centering
		\includegraphics[width=0.8\textwidth]{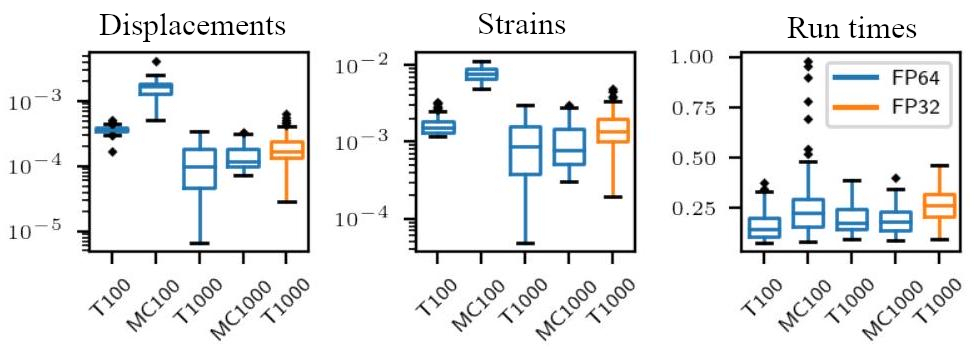}
		\caption{Relative $L_{2}$ errors and run times of DEM for trapezoidal rule (T) and Monte Carlo integration (MC) on 100 and 1000 grid points each.}
		\label{fig:4-10}
	\end{figure}

	\paragraph{Example A}
	We enforced the geometric boundary conditions by the transformation
	\begin{align}
		u(X)=(1+X) z_{\theta}(X)\,\,,
		\label{eq:4-22}
	\end{align}
	where $z_{\theta}(X)$ is the output of the NN. This way,  always zero displacement is calculated at the clamped end ($X = -1$).
	A comparison of the resulting relative errors in displacements and strains as well as the run times is shown in Fig. \ref{fig:4-10}. 
	
	Two integration methods (Monte Carlo integration and trapezoidal rule) are compared, each on different sets of randomly selected support points (100 and 1000, respectively). For the trapezoidal rule, the support points are sorted and the boundary points are explicitly considered. 
	The optimizer is the L-BFGS with learning rate $1.000$. This proves to be very efficient and approaches the solution after only 15 epochs.
	
	The results furthermore illustrate that the use of single floating point accuracy (FP32) leads to only a slight decrease of accuracy, similar as seen with the conventional PINN. However, the run time even increases with FP32, what indicates slower convergence of the optimization procedure. 

\begin{figure}[h]
		\centering
		\begin{subfigure}{0.46\textwidth}
			\includegraphics[width=\textwidth]{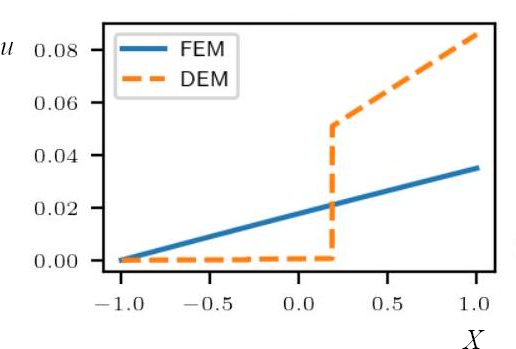}
		\end{subfigure} \hfill
		\begin{subfigure}{0.48\textwidth}
			\includegraphics[width=\textwidth]{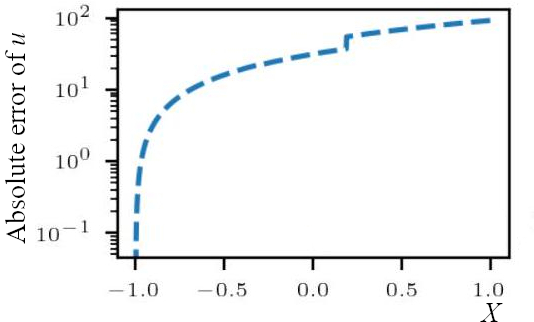}
		\end{subfigure}
		\caption{Displacement results of DEM compared to FEM. Displacements with overfitting (left), Absolute error in displacements (right); rel. $L_2$ error: $1.358 \mathrm{e}+00$ .}	\label{fig:4-12}
	\end{figure}

	\paragraph{Example B}
	The study of Example B reveals a pathology of the DEM that has not appeared in Example A. Its effect can be seen in Fig. \ref{fig:4-12}. This pathology can be attributed to overfitting \cite{Kollmannsberger2021}, since the potential energy, unlike the squared residual, has no regularizing effect. 
	In fact, early stopping significantly reduced the influence of overfitting. Alternatively, overfitting could be avoided by increasing the number of grid points. For 1000 grid points, hardly any overfitting occurred without a need for early stopping.

	\paragraph{Example C}
	The approximation of the energy functional can be done analogously to the 1D example by means of different integration techniques. However, the Monte Carlo integration is the simplest option to implement. 
	The incorporation of the boundary conditions, network architecture and optimizer algorithm as well as the load case are chosen similarly as for the DCM (Section \ref{par:DCM-ProblemC}).
	
\begin{figure}[h!]
		\centering
		\begin{subfigure}{0.49\textwidth}
			\centering
			\includegraphics[width=0.7\textwidth]{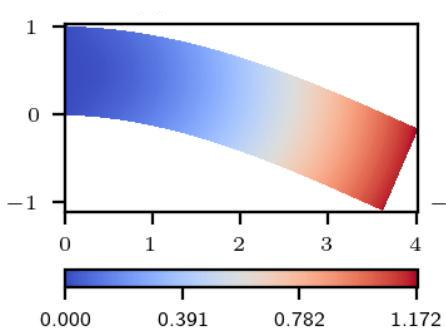}
			\caption{Displacements}
		\end{subfigure}
		\begin{subfigure}{0.49\textwidth}
			\centering
			\includegraphics[width=0.7\textwidth]{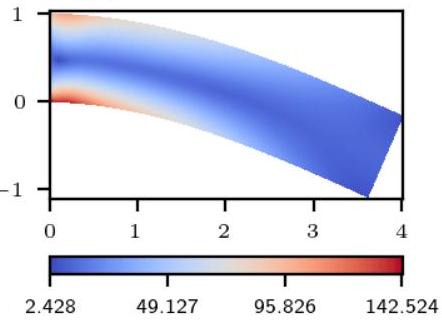}
			\caption{Equivalent stresses}
		\end{subfigure}
		\caption{Calculated displacements and equivalent stresses for the vertical load case.}	\label{fig:4-15}
	\end{figure}

\begin{figure}[h!]
		\centering
		\begin{subfigure}{0.49\textwidth}
			\centering
			\includegraphics[width=0.7\textwidth]{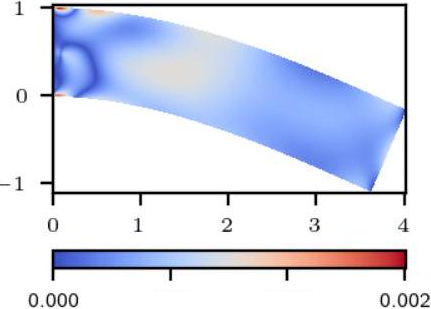}
			\caption{Error in displacements}
		\end{subfigure}
		\begin{subfigure}{0.49\textwidth}
			\centering
			\includegraphics[width=0.7\textwidth]{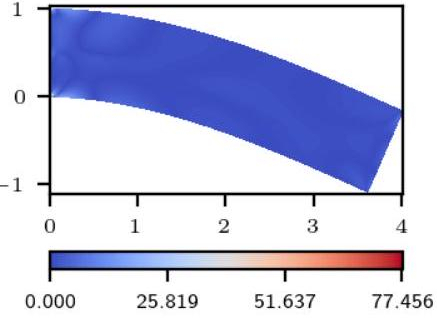}
			\caption{Error in equivalent stresses}
		\end{subfigure}
		\caption{Absolute error in displacements and equivalent stresses for the vertical load case.}	\label{fig:4-16}
	\end{figure}

	10 000 collocation points within the body bulk are selected. The external energy is evaluated by using 200 random points on the right edge of the plate. 
	The results for load case C1 and error are presented in Figs. \ref{fig:4-15} and \ref{fig:4-16}, respectively.
	
	The relative $L_{2}$ error in the displacements is only 0.002 and in the equivalent stress 0.053. Fig. \ref{fig:4-16} shows that the error in the equivalent stresses is concentrated at the restraint. The stress peaks at the critical points are not completely resolved by the NN. 
	
	The integration method does not cause this error, which is confirmed by a second calculation that uses trapezoidal rule. The relative $L_{2}$ errors in the case are 0.003 for displacements and 0.039 for equivalent stresses. Again, an equidistant grid with 10 000 collocation points is used. No significant improvement is obtained by the more accurate integration procedure, so the influence must be considered small.
	
\begin{figure}[h!]
		\centering
		\begin{subfigure}{0.49\textwidth}
			\centering
			\includegraphics[width=0.7\textwidth]{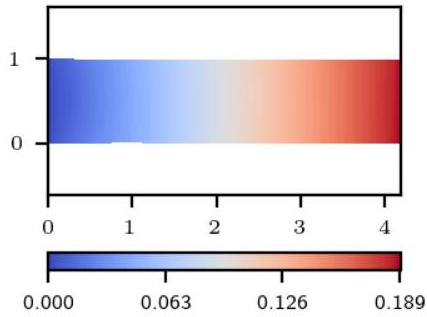}
			\caption{Displacements}
		\end{subfigure}
		\begin{subfigure}{0.49\textwidth}
			\centering
			\includegraphics[width=0.7\textwidth]{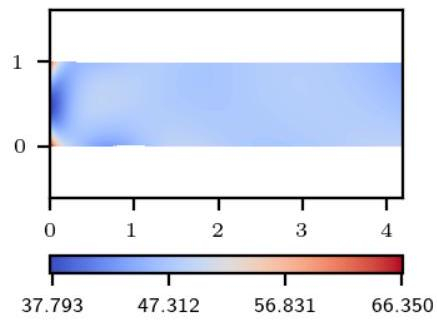}
			\caption{Equivalent stresses}
		\end{subfigure}
		\caption{Calculated displacements and equivalent stresses for uniaxial tension.}	\label{fig:4-20}
	\end{figure}

	For load case C2, the same procedure with MC integration is carried out. The results and errors are presented in Figs. \ref{fig:4-20} and \ref{fig:4-16}, respectively. The relative $L_{2}$ errors are 0.005 and 0.019 in this case. 
	
	In conclusion, the NN is able to approximate the character of the solution of the BVP. However, relatively large errors are found at the restraint. According to \cite{FuhgBouklas2022}, the same problems arise for a PINN trained with the squared residual. The resolution of fine features of the stress and displacement fields seem still to be a challenge for future work.
	
	\begin{figure}[h!]
		\centering
		\begin{subfigure}{0.49\textwidth}
			\centering
			\includegraphics[width=0.7\textwidth]{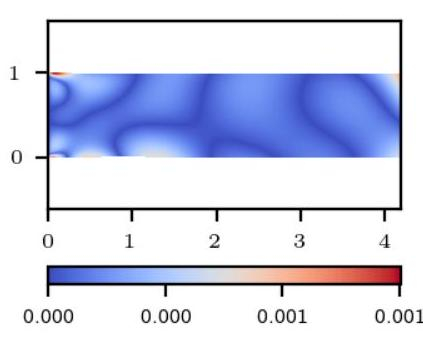}
			\caption{Error in Displacements}
		\end{subfigure}
		\begin{subfigure}{0.49\textwidth}
			\centering
			\includegraphics[width=0.7\textwidth]{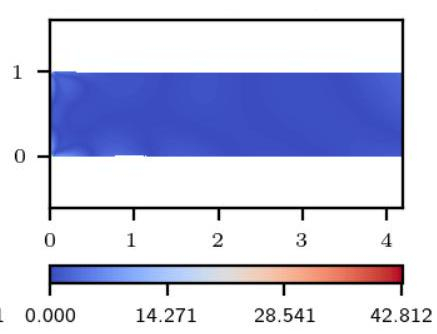}
			\caption{Error in Equivalent stresses}
		\end{subfigure}
		\caption{Absolute error in the calculated displacements and equivalent stresses for uniaxial tension.}	\label{fig:4-21}
	\end{figure}

	\subsubsection{Transfer learning (TF)}
	One possibility to enhance the performance of Neural FEM is by means of transfer learning e.g. in case of a varying Neumann boundary condition. This is illustrated on Example B2, where $\pi_{2}$ is changed by only a small amount in each iteration. Then the NN trained on the previous $\pi_{2}$ value is already a good approximation for its subsequent value. Hence, the NN parameter values can be copied to the new NN to reduce the number of required learning epochs. 
	
\begin{figure}[h!]
		\centering
		\includegraphics[width=0.4\textwidth]{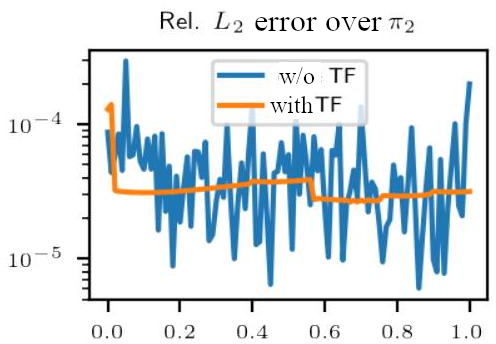}
		\caption{Effect of transfer learning (TF). Comparison of $L_{2}$ errors over the parameter space.}
		\label{fig:4-34}
	\end{figure}

	Applied to Example B2 (linear elastic material), the average run time could be reduced by a factor of four. Similar time savings have been documented in \cite{Abueidda2021} in the study of plastic deformations. The relative $L_{2}$ errors are shown in Fig. \ref{fig:4-34}. The average $L_{2}$ error is 3.145 $\cdot 10^{-5}$, which is about an order of magnitude smaller than the error with DeepONet. The training duration is reduced from $23 \mathrm{~s}$ to about $5 \mathrm{~s}$.
	
	\subsubsection{Initialization of a conventional FE solver}
	As indicated in the literature and the results above, relative $L_{2}$ errors of approx. $10^{-4}$ are usually achieved. The computed solution could then be submitted as initialization to a traditional FEM solver in order to improve the accuracy. 
	
	Applied to Example A, a FEniCS calculation, that is conventionally initialized with all displacements as zero, runs for four iterations.
	With the solution of the DEM with trapezoidal rule and 1000 collocation points as initialization values for the FEM solver, the calculation is accelerated by a factor of two, only half of the iterations until convergence are needed.
	
	Thus, Neural FEM results can be employed as a potential way to speed up an FEM simulation in settings where the Neural FEM is not yet able to completely replace the FE simulation.

	\subsection{Neural Operators}
	The operator methods are examined on example Example B2 (Section \ref{par:Problem-B2}) -- a tensile bar with clamping restraint at the left side and a free end at the right side.
	In this simple test case, analytical solutions are available to calculate the error for the parameterization of the Neumann boundary condition in case of fixed force density $f(X) \equiv 1$. Combinations of both varying the force field and the boundary conditions have not been carried out in the present work.
	
\begin{figure}[h!]
		\centering
		\begin{subfigure}{0.49\textwidth}
			\includegraphics[width=0.8\textwidth]{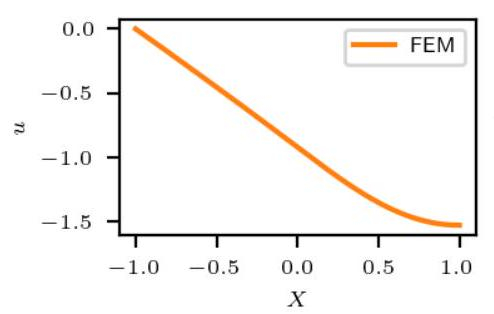}
			\caption{Displacements}
		\end{subfigure}
		\begin{subfigure}{0.49\textwidth}
			\includegraphics[width=0.8\textwidth]{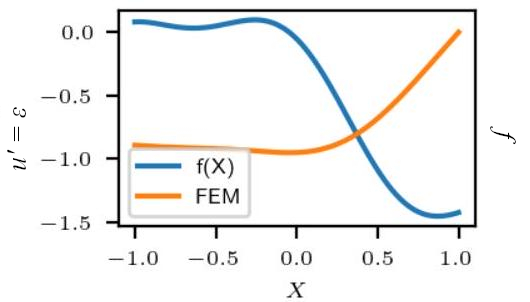}
			\caption{Stresses $P=u^{\prime}$ and force density $f$}
		\end{subfigure}
		\caption{FEM solution for a random realization of the force density $f$.}	
		\label{fig:4-24}
	\end{figure}

	Training and test data for the varying force field are generated by means of Gaussian Processes with squared exponential covariance (correlation length $l=0.100$). 
	The free end of the bar at the right side yields the parameter $\pi_{2} = 0$.
	For training, 1000 data sets and for the tests 100 data sets have been generated with FEniCS on an equidistant grid for $X \in [-1,1]$ with 1024 grid points and quadratic shape functions. The relative $L_{2}$ errors of the FEM simulation are several orders of magnitude smaller than expected from the NN methods (displacements approx. $10^{-10}$; strains approx. $ 10^{-7}$) and should not influence the survey. One of the resulting data sets is illustrated in Fig. \ref{fig:4-24}.

	\subsubsection{DeepONet and PIDeepONet}
	\paragraph{Numerical setup}
	For the DeepONet, the data sets need to be preprocessed since the Neural Operator methods work with $P$ random collocation points that change between the evaluations of the loss function,
	whereas the reference solutions are produced be FEM on a fix mesh with $m$ equidistant points. Hence, the realizations of the force fields are projected onto the FE mesh. Then, the FE results are interpolated and evaluated at the random collocation points.
	The parametrization of Neumann boundary condition has only been investigated with DeepONet, where $\pi_{2} \sim U[0,1]$ is assumed.
	
	The source codes for DeepONet and PIDeepONet have both been published on Github \cite{lu2021learning, wang2021learning}. The architectures for the subnets were specified as $[20,100,100]$ for the branch net and $[1,100,100]$ for the trunk net. The branch is set up with layer width $m=20$. The chosen activation function are ReLU in DeepONet and Tanh in PIDeepONet. A similar architecture has also been suggested in \cite{Lu2021, Wang2021b}. The L-BFGS optimizer with linesearch (strong Wolfe condition) and learning rate $1.0$ is applied as suggested in \cite{Krishnapriyan2021}, for 120 epochs. However, L-BFGS is very memory consuming, so $historysize =50$ (default: 100) is set. 
	
	1000 load cases are used for training and 100 for testing.
	In order to reduce the training effort, from the 1024 grid points only 8 are randomly chosen per load case. Hence, the training data set size reduces to 8000 data points. The error reduction of considering more grid points per load case quickly saturates so that this reduction is admissible. Approximately 50 random points can be estimated as the saturation limit for 1000 training data sets \cite{Lu2021}. Training is conducted for both methods, Full-Batch and Mini-Batch (batch size: 1000) .
	
	Since PIDeepONet includes the whole information about the PPDE (similar to the PINN models) in the loss function, no reference solutions by means of FEM are necessary. 
	In exchange, the loss functions needs to be constructed anew for each PDE.
	
	For the DeepONet, additionally an alternative loss function is investigated. It employs the relative $L_{2}$ error instead of the Mean Square Error  -- MSE (Eq. \ref{eq:MSE}). 
	
\begin{figure}
		\centering
		\begin{subfigure}{0.49\textwidth}
			\centering
			\includegraphics[width=0.8\textwidth]{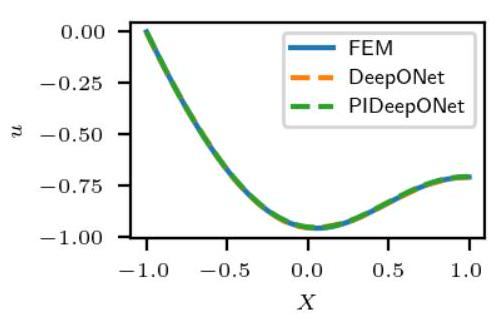}
			\caption{Displacements}
		\end{subfigure}
		\begin{subfigure}{0.49\textwidth}
			\centering
			\includegraphics[width=0.8\textwidth]{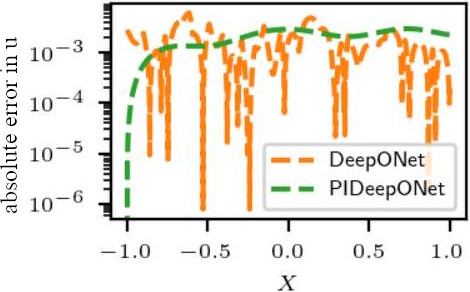}
			\caption{Absolute error for displacements}
		\end{subfigure}
		\begin{subfigure}{0.49\textwidth}
			\centering
			\includegraphics[width=0.8\textwidth]{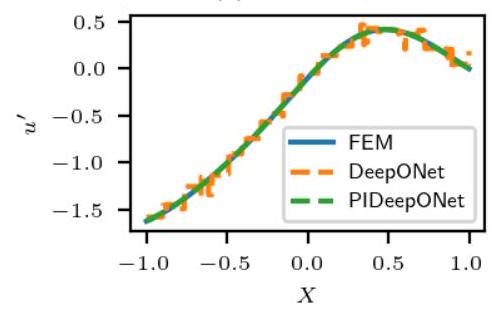}
			\caption{Strains}
		\end{subfigure}
		\begin{subfigure}{0.49\textwidth}
			\centering
			\includegraphics[width=0.8\textwidth]{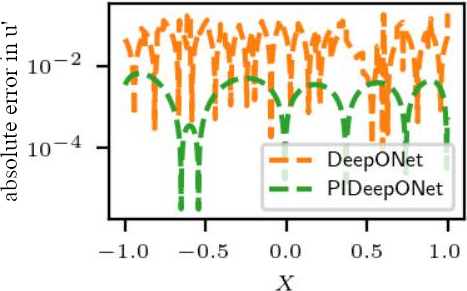}
			\caption{Absolute error for strains}
		\end{subfigure}
		\caption{Displacements and strains with corresponding errors for an exemplary element of the test data set with\\ DeepONet.}	
		\label{fig:4-30}
	\end{figure}

	\paragraph{Results}
	Representative results for the displacement $u$ and strains $u'$ over the bar length obtained with DeepONet and PIDeepONet are shown in Fig. \ref{fig:4-30}. 
	Both methods match the displacements relatively well, but DeepONet has visible deviations in the strains. In particular, the non-smooth curve of the strain, which is the spatial derivative of the displacement, can be attributed to the ReLU activation function, which has a discontinuous derivative.
		
	The errors for the displacements and strains are shown in Table \ref{tab:4-7}.
	The usage of the residual in the empirical risk in PIDeepONet improves the accuracy in the strains about one order of magnitude compared to DeepONet.

	The effect of floating point accuracy on $\epsilon_{\mathrm{rel}}$ is small, similar as seen for Neural FEM.	
	Overall, training with Full-Batch on FP32 performs best.
		
	The runtimes of models, each trained with Full-Batch and Mini-Batch (batch size 1000), are compared in Table \ref{tab:4-9}. The run times for the DeepONet are significantly lower than for the FNO with comparable accuracies. Using the relative $L_{2}$ error instead of the MSE reduces the convergence rate, requiring more iterations and increasing the run time. The difference between the mean and median is reduced, but no significant effect on the error is found.
		
	\begin{table}
		\centering
		\begin{tabular}{lllll}
			\hline
			& Mean(FP32) & Median(FP32) & Mean(FP64) & Median(FP64) \\
			\hline
			DeepONet(MSE), displacements & $5.070 \cdot 10^{-3}$ & $3.144 \cdot 10^{-3}$ & $5.076 \cdot 10^{-3}$ & $3.023 \cdot 10^{-3}$ \\
			DeepONet(L2), displacements & $-$ & $-$ & $2.109 \cdot 10^{-3}$ & $1.329 \cdot 10^{-3}$ \\
			PIDeepONet, displacements & $3.268 \cdot 10^{-3}$ & $2.008 \cdot 10^{-3}$ & $2.450 \cdot 10^{-3}$ & $1.343 \cdot 10^{-3}$ \\
			\hline
			DeepONet(MSE), strains & $4.022 \cdot 10^{-2}$ & $2.831 \cdot 10^{-2}$ & $4.023 \cdot 10^{-2}$ & $2.833 \cdot 10^{-2}$ \\
			DeepONet(L2), strains & $-$ & $-$ & $5.393 \cdot 10^{-2}$ & $4.492 \cdot 10^{-2}$ \\
			PIDeepONet, strains & $3.055 \cdot 10^{-3}$ & $2.394 \cdot 10^{-3}$ & $2.561 \cdot 10^{-3}$ & $1.936 \cdot 10^{-3}$ \\
			\hline
		\end{tabular}
		\caption{Mean error values over the test data set with DeepONet and PIDeepONet.}\label{tab:4-7}
	\end{table}
	
	\begin{table}
		\centering
		\begin{tabular}{lllll}
			\hline
			& Full (FP64) & Mini (FP64) & Full (FP32) & Mini (FP32) \\
			\hline
			DeepONet(MSE) & $82 \mathrm{~s}$ & $188 \mathrm{~s}$ & $70 \mathrm{~s}$ & $120 \mathrm{~s}$ \\
			DeepONet(L2) & $112 \mathrm{~s}$ & $200 \mathrm{~s}$ & $-$ & $-$ \\
			PIDeepONet & $933 \mathrm{~s}$ & $951 \mathrm{~s}$ & $607 \mathrm{~s}$ & $-$ \\
			\hline
		\end{tabular}
		\caption{Run times of the optimization procedure with Full-Batch and Mini-Batch (1000 elements), respectively.} \label{tab:4-9}
	\end{table}

\begin{figure}
		\centering
		\includegraphics[width=0.8\textwidth]{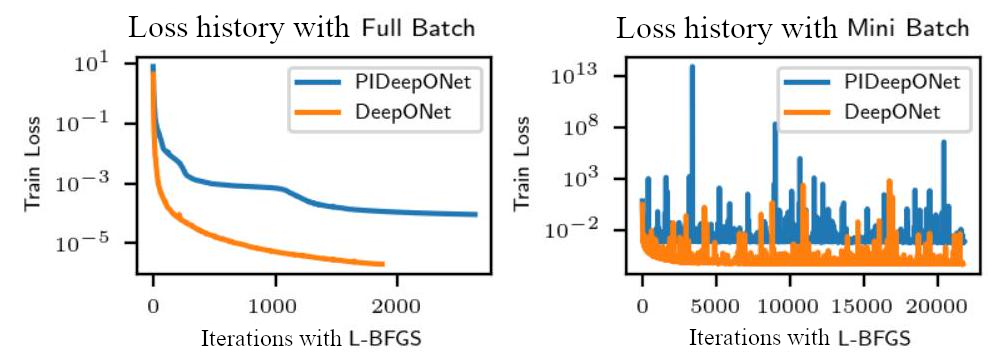}
		\caption{Training loss histories after optimizing PIDeepONets with Full-Batch and Mini-Batch.}
		\label{fig:4-32}
	\end{figure}

	Fig. \ref{fig:4-32} shows the loss histories from the optimization with the DeepONet and PIDeepONet, respectively. Overall, the relative $L_2$ error on the test data set is smaller for Full Batch training.
	The difference in resulting accuracy between the two methods can be attributed to the training error alone.
	PIDeepONet converges significantly slower and yields worse accuracy than DeepONet. With Full-Batches, PIDeepONet even converges to a local instead of the global minimum. 
	The poor convergence of the PIDeepONet demonstrates the significantly more complex optimization task, where the DeepONet makes use of the explicitly obtained FEM results as training data sets.

	\paragraph{Potential Energy in the loss function}
	The results of Neural FEM (Section \ref{subsubsec:cPINN-results}) and \cite{Karumuri2020, NabianMeidani2019} suggest, that replacing the squared residual in the loss function of PIDeepONet by the potential energy can make the optimization problem easier to solve. Hence, such a method should be more robust and efficient and make training feasible even where training of PINNs fails. However, with 100 random realizations of the force field and 15 collocation points per realization as suggested in \cite{Karumuri2020}, this method does not converge for Example B2.

	\paragraph{Influence of intialization}
	The default initialization by PyTorch sets the weights and bias by randomly sampling from a uniform distribution. For $\boldsymbol{W} \in \mathbb{R}^{N_{k} \times N_{k-1}}$ and $\boldsymbol{b} \in \mathbb{R}^{N_{k}}$, it holds:
	
	\begin{align}
		W_{i j}, b_{i} \sim U[-\sqrt{k}, \sqrt{k}] \quad \text { with } \quad k=\frac{1}{N_{k-1}},
		\label{eq:4-37}
	\end{align}
	
	where $N_{k}$ denotes the width of the $k$-th layer. \citeauthor*{Glorot2010} \cite{Glorot2010} and \citeauthor*{Wang2021b} \cite{Wang2021b} suggest that the convergence of the NN can be accelerated by the Glorot initialization. Let $N\left[\mu, \sigma^{2}\right]$ be a normal distribution with mean $\mu = 0$ and variance $\sigma^{2}$. This yields
	\begin{align}
		b_{i}=0 \quad \text { and } \quad W_{i j} \sim N\left[0, \sigma^{2}\right] \quad \text { with } \quad \sigma=\sqrt{\frac{2}{N_{k}+N_{k-1}}}
		\label{eq:4-38}
	\end{align}
	for the parameters.

	The errors of the models on single precision with Full-Batch optimization and L-BFGS are shown in Table \ref{tab:4-10}. The error of DeepONet for the displacements becomes only slightly smaller. Also, no large difference in run times was observed. The Glorot initialization was developed mainly for deep learning applications and does not have much impact on the shallow networks (with one hidden layer) used here.
	
	\begin{table}
		\centering
		\begin{tabular}{llll}
			\hline
			& displacements & strains & running time \\
			\hline
			DeepONet & $1.629 \cdot 10^{-3}$ & $3.594 \cdot 10^{-2}$ & $64 \mathrm{~s}$ \\
			PIDeepONet & $5.338 \cdot 10^{-3}$ & $4.501 \cdot 10^{-3}$ & $601 \mathrm{~s}$ \\
			\hline
		\end{tabular}
		\caption{Mean error values over the test data set, with Glorot initialization.} \label{tab:4-10}
	\end{table}

	\paragraph{Neumann boundary parameterization}
	The (PI-)DeepONet and the FNO have been specifically designed for approximating mappings between function spaces. One advantage of the DeepONet over the FNO is that it is very easy to apply to arbitrary parameterizations. For example, the PIDeepONet can be used to parameterize the Neumann boundary. Let again be given the dimensionless problem
	
	\begin{align}
		-u^{\prime \prime}(X)=1 \quad \text { with } \quad u(-1)=0, \,\, u^{\prime}(1)=\pi_{2} \quad .
		\label{eq:4-39}
	\end{align}
	
	The analytical solution is given by
	
	\begin{align}
		u(X)=\frac{3}{2}+X-\frac{1}{2} X^{2}+\pi_{2}(X+1) \quad .
		\label{eq:4-40}
	\end{align}
	
	In the following, only the scalar variable $\pi_{2}$ needs to be varied. The solution operator $G: I \rightarrow \mathcal{S}$ is sought, where $I=[0,1]$ is fixed and $\mathcal{S}$ denotes the space of admissible deformations.
	
	For each of the subnetworks, a hidden layer with 50 neurons is used. Their architectures are thus given by $[1,50,50]$. Tanh is used as the activation function according to the experience in neural FEM. The NN is trained over 40 epochs with L-BFGS on 10 000 training data set entries, and 1000 validation data set entries. 
	
\begin{figure}
		\centering
		\includegraphics[width=0.55\textwidth]{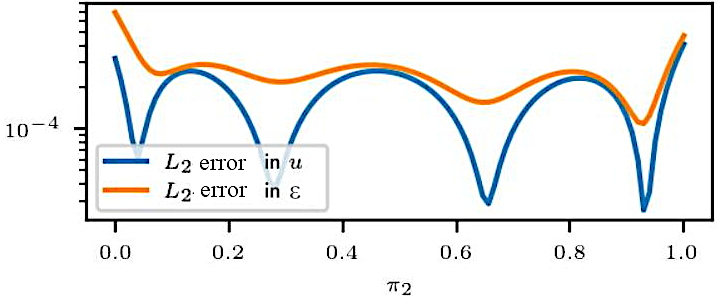}
		\caption{Distribution of the relative $L_{2}$ error over the parameter space for displacements $u$ and strains $\varepsilon$.}
		\label{fig:4-33}
	\end{figure}

	The data set is built by selecting a single random collocation point for each of the 100 realizations of $\pi_{2}$ ($P=1$) and 1000 grid points each on the interval $[-1,1]$.
	No early stopping is used. The relative $L_{2}$ error over the whole interval $I$ is shown in Fig. \ref{fig:4-33}.

	The training of the mesh took approximately $118 \mathrm{~s}$. The mean relative $L_{2}$ error for the displacements is $1.695 \cdot 10^{-4}$  and $2.463 \cdot 10^{-4}$ for the strains. For the calculation of the complete test data set, the PIDeepONet took $0.033 \mathrm{~s}$. This highlights the difference between the short run time of the inference and the large computational effort for the training.
	
	With L-BFGS and 1000 collocation points (to minimize the risk of overfitting) the training of the DEM on 100 realizations took about 23 s. The higher training effort of the operator model is profitable only for about 500 realizations of $\pi_{2}$ and more.

	\subsubsection{FNO and PINO}
	The FNO architecture proposed in \cite{Li2020b} is applied for Example B2. Here, the hyperparameter representing the hidden layer width $d_v = 64$ (Section \ref{subsec:FNO}) results in 549 569 NN parameters.
	Secondly, a smaller architecture with $d_v = 12$ is set up which results in 20 885 NN parameters. This is comparable to the DeepONet architecture with 22 500 parameters.
	
	Padding was considered as suggested in \cite{Li2020b}, since the considered example has non-periodic boundary conditions in the input functions.
	
	Gaussian Error Linear Unit (GELU) is used as the activation function, 
	\begin{align}
		\operatorname{GELU}(x) = x \Phi(x)=\frac{x}{2}\left[1+\operatorname{erf}\left(\frac{x}{\sqrt{2}}\right)\right],
		\label{eq:4-34}
	\end{align}
	where $\Phi(x)$ denotes the standard normal distribution. 
	
	Adam with decreasing learning rate (initial value 0.001, reduction factor 0.5 every 50 epochs) and weight decay $\lambda = 10^{-4}$ is used as optimizer. The other optimizer parameters are kept as the PyTorch default values. The relative $L_{2}$ error is used for the loss function.
	
	Similar to the DeepONet, a training data set with 1000 entries and a test data set with 100 entries is used, 500 training epochs are carried out.
	
	In the present work, the spatial derivatives in the loss value for the elastic strain energy and body forces are approximated by a second order central difference method instead of employing the autograd feature. This can significantly reduce the computational effort, since the number of parameters in the NN is usually much greater than the number of grid points. Exact derivation methods are discussed in more detail in \cite{Li2022}.

\begin{figure}[h]
		\centering
		\begin{subfigure}{0.49\textwidth}
			\centering
			\includegraphics[width=0.8\textwidth]{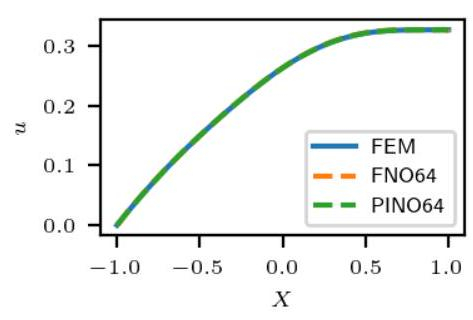}
			\caption{Displacements}
		\end{subfigure}
		\begin{subfigure}{0.49\textwidth}
			\centering
			\includegraphics[width=0.8\textwidth]{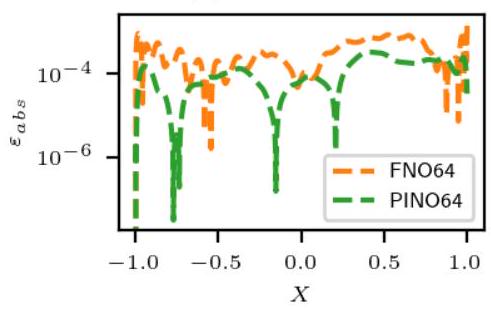}
			\caption{Absolute error}
		\end{subfigure}
		\caption{Displacements calculated with FNO and PINO (for an exemplary load case of the test data set).}	
		\label{fig:4-25}
	\end{figure}

\begin{figure}[h]
		\centering
		\begin{subfigure}{0.49\textwidth}
			\centering
			\includegraphics[width=0.8\textwidth]{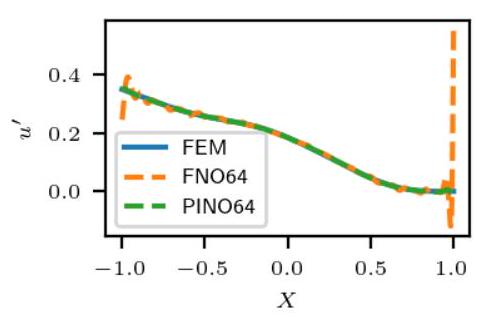}
			\caption{Strains}
		\end{subfigure}
		\begin{subfigure}{0.49\textwidth}
			\centering
			\includegraphics[width=0.8\textwidth]{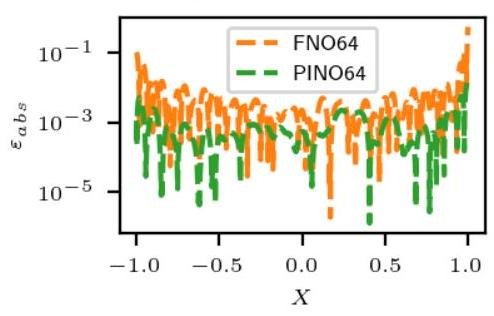}
			\caption{Absolute error}
		\end{subfigure}
		\caption{Strains calculated with FNO and PINO (for an exemplary load case of the test data set).}	
		\label{fig:4-26}
	\end{figure}

	The results for an exemplary test load case are shown in Figs. \ref{fig:4-25} and \ref{fig:4-26}. The absolute errors in the displacements computed by FNO and PINO are relatively similar, but the strains at the endpoints of the bar as computed by the pure FNO show significant errors, rendering the solution practically unusable. The mean errors in the displacements and strains on the whole test data set are shown in Table \ref{tab:4-4}.
	The inclusion of potential energy does not yield a significant effect on the accuracy for the displacements. The unphysical oscillations of the calculated strains at the edges of the computational domain are eliminated (Fig. \ref{fig:4-26}).
	The only drawback of PINO is the discretization dependence of the numerical derivative. After training, the error is no longer constant over different discretization levels which is analyzed in Section \ref{subsubsec:Zero-shot}.

	\begin{table}[h]
		\centering
		\begin{tabular}{lllll}
			\hline
			& Mean ($d_v = 64$) & Median ($d_v = 64$) & Mean ($d_v = 12$) & Median ($d_v = 12$) \\
			\hline
			FNO, displacements & $2.438 \cdot 10^{-3}$ & $1.447 \cdot 10^{-3}$ & $1.102 \cdot 10^{-2}$ & $6.428 \cdot 10^{-3}$ \\
			PINO, displacements & $2.688 \cdot 10^{-3}$ & $1.528 \cdot 10^{-3}$ & $7.011 \cdot 10^{-3}$ & $4.436 \cdot 10^{-3}$ \\
			\hline
			FNO, strains & $1.145 \cdot 10^{-1}$ & $9.278 \cdot 10^{-2}$ & $3.488 \cdot 10^{-1}$ & $3.119 \cdot 10^{-1}$ \\
			PINO, strains & $1.382 \cdot 10^{-2}$ & $1.053 \cdot 10^{-2}$ & $4.941 \cdot 10^{-2}$ & $3.835 \cdot 10^{-2}$ \\
			\hline
		\end{tabular}
		\caption{Mean and median values for errors in displacements and strains from FNO and PINO with $d_{v}=64$ and $d_{v}=12$.} \label{tab:4-4}
	\end{table}
	
\begin{figure}[h]
		\centering
		\begin{subfigure}{0.49\textwidth}
			\centering
			\includegraphics[width=0.8\textwidth]{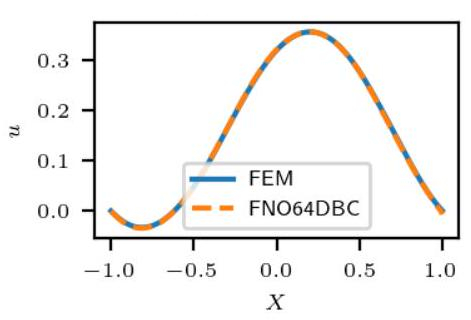}
			\caption{Displacements}
		\end{subfigure}
		\begin{subfigure}{0.49\textwidth}
			\centering
			\includegraphics[width=0.8\textwidth]{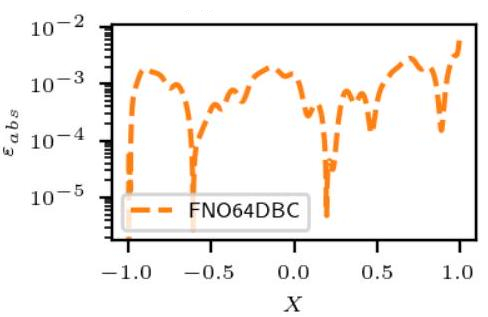}
			\caption{Absolute error for displacements}
		\end{subfigure}
		\begin{subfigure}{0.49\textwidth}
			\centering
			\includegraphics[width=0.8\textwidth]{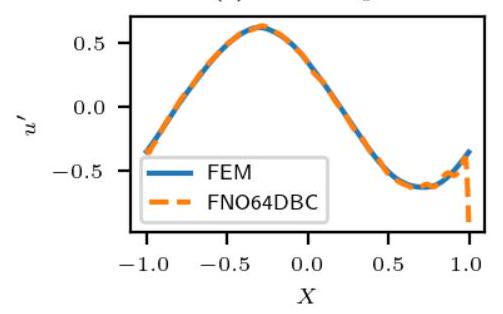}
			\caption{Strains}
		\end{subfigure}
		\begin{subfigure}{0.49\textwidth}
			\centering
			\includegraphics[width=0.8\textwidth]{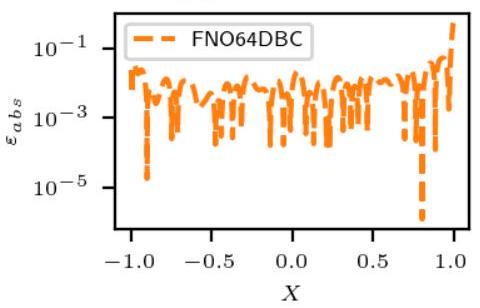}
			\caption{Absolute error for strains}
		\end{subfigure}
		\caption{Displacements, strains and errors calculated with FNO for the Burgers data set (load case with highest $L_2$ error within the test data set).}	
			\label{fig:4-27}
	\end{figure}

	The errors are concentrated at the edge, which indicates that the choice of boundary conditions for the FNO might be unfavorable, although this approach should be able to map non-periodic boundary conditions when padding data arrays with zeros. For comparison, a new data set with periodic force fields and a bar clamped on both sides has been modeled. For this purpose, the data set for the 1D-Burgers problem is assumed \cite{Li2020b}. The periodic initial conditions are interpreted as force fields. The results that yield the largest relative $L_{2}$ error among the test data set are shown in Fig. \ref{fig:4-27} for the displacements and strains.

	The reason for the poor agreement at the boundaries cannot be attributed solely to the non-periodic boundary conditions of the input functions, although for the periodic boundary conditions, the median of the $\epsilon_{\text {rel}}$ error is lower for the periodic boundary conditions ($2.396 \cdot 10^{-2}$) than for the non-periodic ones
	($9.278 \cdot 10^{-2}$). Large deviations at the boundaries of the periodic domain are still present (Fig. \ref{fig:4-27}). These errors can be reduced by regularization (for example by means of an energy functional), as observed with the Neural FEM.

	The average run times per epoch for computations on CPU and GPU are shown in Table \ref{tab:4-5}. The midrange mobile graphics in use accelerated the calculation by factor 4 for the 32 Bit floating point accuracy. For a comparable acceleration for FP64, instead of consumer graphics cards specific High Performance Computing (HPC) accelerator cards are necessary.
	We skip a calculation of the $\epsilon_{\mathrm{rel}}$ errors with double precision because no large influence on the total error can be expected, according to the results of the Neural FEM.

	\begin{table}
		\centering
		\begin{tabular}{lllll}
			\hline
			& CPU ($d_v = 64$) & GPU ($d_v = 64$) & CPU ($d_v = 12$) & GPU ($d_v = 12$)\\
			\hline
			FNO (FP32) & $9.754 \mathrm{~s}$ & $2.447 \mathrm{~s}$ & $2.737 \mathrm{~s}$ & $0.765 \mathrm{~s}$ \\
			PINO (FP32) & $9.926 \mathrm{~s}$ & $2.491 \mathrm{~s}$ & $2.874 \mathrm{~s}$ & $0.837 \mathrm{~s}$ \\
			\hline
			FNO (FP64) & $18.120 \mathrm{~s}$ & $14.694 \mathrm{~s}$ & $5.163 \mathrm{~s}$ & $4.458 \mathrm{~s}$ \\
			PINO (FP64) & $17.041 \mathrm{~s}$ & $14.763 \mathrm{~s}$ & $5.076 \mathrm{~s}$ & $4.491 \mathrm{~s}$ \\
			\hline
		\end{tabular}
		\caption{Average runtimes per epoch of FNO and PINO on FP32 and FP64 with $d_{v}=64$ and $d_{v}=12$.}\label{tab:4-5}
	\end{table}

	\subsubsection{Zero-shot super resolution}	\label{subsubsec:Zero-shot}
\begin{figure}
		\centering
		\includegraphics[width=0.55\textwidth]{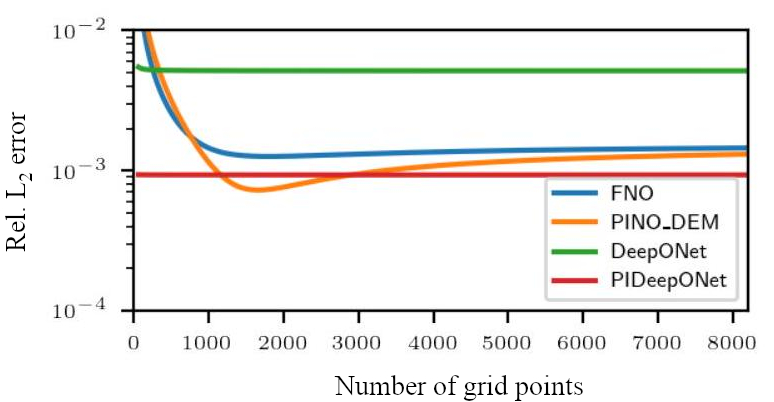}
		\caption{Relative $L_2$ error on a single instance of the test data set with varying discretization used for NN evaluation.}	
		\label{fig:4-35}
	\end{figure}

	A main feature of neural operators is the consistency of the numerical error over different discretization levels. This is manifested by a nearly constant progression of the error across the discretization level, as shown in Fig. \ref{fig:4-35}. Therefore, zero-shot super resolution becomes possible, which means that the NN can be evaluated on a finer grid than the one that used for training. In the present work, the NNs are trained on a dataset based on FEM solutions with a grid with 1024 nodes, but can also be evaluated on finer discretization levels with the same error. The only exception is the PINO which uses a finite difference method in the optimization process in the current contribution. 
	The reference solution for the finer discretization task is obtained from an FEM analysis with 8192 node points. This data set is cubically interpolated to all other discretizations for comparison with the results of the NNs.
		
	\begin{table}
		\centering
		\begin{tabular}{lllll}
			\hline
			Model & Error in $u$ & Error in $u_{,X}$ & Run time on CPU in s & Run time on GPU in s\\
			\hline
			FNO & $2.438 \cdot 10^{-13}$ & $1.145 \cdot 10^{-1}$ & $81.289$ (estimated) & $20.392$\\
			PINO & $2.688 \cdot 10^{-13}$ & $1.382 \cdot 10^{-2}$ & $82.713$ (estimated) & $20.755$\\
			DeepONet & $5.070 \cdot 10^{-13}$ & $4.022 \cdot 10^{-2}$ & $1.167$ & --\\
			PIDeepONet & $2.783 \cdot 10^{-13}$ & $2.800 \cdot 10^{-3}$ & $10.117$ & --\\
			\hline
		\end{tabular}
		\caption{Comparison of the $\epsilon_{rel}$ errors in the test data set with FP32.} \label{tab:5-1}
	\end{table}
	
	\section{Summary and Outlook} \label{sec:SumOut}
	In this work, different NN methods have been analyzed and applied to examples from elastostatics. Specifically, a 1D tensile bar with a hyperelastic material (Example A, Section \ref{subsubsec:ProblemA}) and with a linear elastic material (Example B, Section \ref{subsubsec:ProblemB}) has been investigated. Moreover, a plate made of a Neo-Hookean material has been analyzed for two load cases, vertical loading and uniaxial tension (Example C, Section \ref{subsubsec:4-2-4}).
	
\begin{figure}
		\centering
		\includegraphics[width=0.8\textwidth]{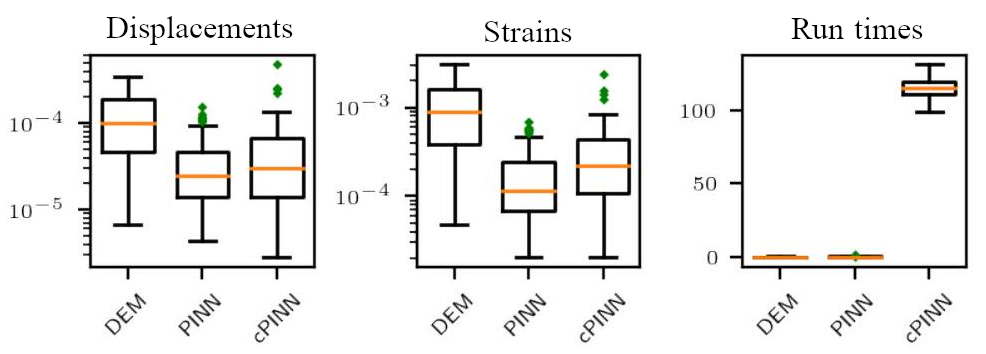}
		\caption{Relative $L_{2}$ errors and run times for Example A with the Neural FEM approaches.}
		\label{fig:4-22}
	\end{figure}

	\paragraph{Physics Informed Neural Networks (PINN)}
	In the basic form of classical PINN \cite{Raissi2019}, the empirical risk is built from the squared residuals of the differential operators. In various works \cite{Krishnapriyan2021, Wang2021a, Liu2021}, it has been shown that such a PINN is difficult or impossible to train even for simple examples, so alternative forms of regularization have been developed. The present work particularly studies the DEM, based on the principle of minimal potential energy, and the cPINN, based on the game-theory. The results for Example A using these three approaches (PINN, DEM, cPINN) are compared in Fig. \ref{fig:4-22}. The average accuracies are relatively similar, but the comparatively long run time of the cPINN is disadvantageous.
		
	According to \cite{Zeng2022}, the relative errors can be reduced by up to 2 orders of magnitude by the cPINN, which could not be demonstrated with Example A in this work. 
	The training of cPINN and DEM should converge in more cases than pure PINN, i.e. it is more robust, as demonstrated with Example C.
	Overall, the PINN performes best in Example A. However, the latter is not suitable to show the training pathologies of the PINN. 
	Those pathologies were demonstrated only on Example C, where the training of the PINN fails, but the DEM can be applied successfully. The error measured in the $L_{2}$ norm is 
	relatively small. But the absolute errors of the equivalent stresses for the 2D plate in the vertical load case show deviations of up to $77 \mathrm{~Nm}^{-2}$ at the restraint, for a maximum stress of about $142 \mathrm{~N}\mathrm{m}^{-2}$. 
			
	Our results with Example C underline the suggestion from \cite{FuhgBouklas2022} that PINN and DEM as well as PIDeepONet in their present form are not able to resolve stress concentrations. Further work is necessary to find and analyze alternative approaches with improved accuracy and applicability for classical tasks in solid mechanics like the investigation of critical areas in strength analysis.
	
	\subparagraph{DEM}
	In DEM, the convergence order of the integration method does not seem to be significant after reaching a certain limit of accuracy. 
	However, DEM holds the risk of overfitting, which must be accounted for by early stopping or a sufficient number of collocation points (support points). This topic has not been addressed in the literature up to now.
	
	An advantage from the use of the potential energy is the reduction of the order of differentiation, which also decreases the numerical effort. The run time is about a factor of six lower compared to the PINN. 
	
	All studies show an intense dependency of the result from the initialization of the NN parameters. In extreme cases, the optimizations converges towards different functions. The relative $L_2$ error of the DEM with trapezoidal rule and 1000 collocation points results in the whole range from $4.898 \cdot 10^{-6}$ up to $4.277 \cdot^{-4}$ for displacements and $3.774 \cdot 10^{-5}$ up to $3.701 \cdot 10^{-3}$ for strains. This indicates that the expensive training has to be conducted several times, until an acceptable ML model is found. Further improvement possibilities are Glorot initialization of the NN parameters \cite{Glorot2010} and pretraining \cite{Yagawa2021}.
	
	\paragraph{Neural Operator Methods}
	The Neural Operator methods have been applied to Example B. All models are calculated with single precision floating point since the investigation of DeepONet, PIDeepONet and the Neural FEM did not yield significant effects on the accuracy of the results. Furthermore, the DeepONets have been optimized with Full-Batch training. 
	
	The analysis shows that the DeepONet is significantly faster than the FNO, even if calculated on a GPU. 
	Both FNO and DeepONet can learn the solution operator of the parametric PDE,
	but the achieved accuracies are not sufficient in many cases.

	\paragraph{Outlook}
	According to the results presented, the accuracy of NNs has to be improved for a reliable engineering application in elastostatics.

	One promising approach is a network architecture based on the Neural Attention mechanism, that is claimed to improve the accuracy of PINNs about up to 2 orders of magnitude \cite{Wang2021a}.
	A similar suggestion is made with regard to the Physics-Augmented Learning \cite{Liu2021}.
	The sequential learning and the curriculum learning are two further approaches that might improve the learning ability of PINNs \cite{Krishnapriyan2021}.
	
	In future works, second order methods should be investigated for the optimization of NN parameters during learning \cite{Martens2012, Yao2019}. The example of L-BFGS shows, that the optimization with higher order methods can be immensely accelerated. They also take profit from larger batches, which makes them more efficient in terms of data parallelism.
	
	\clearpage
	\section*{Abbreviations}
	\begin{itemize}
		\item ACGD \dotfill Adaptive Competitive Gradient Descent
		\item ANN \dotfill Artificial Neural Network
		\item BC \dotfill Boundary Conditions
		\item CGD \dotfill Competitive Gradient Descent
		\item cPINN \dotfill competitive Physics Informed Neural Network
		\item DEM \dotfill Direct Energy Method
		\item FCNN \dotfill Fully Connected Neural Network
		\item FEM \dotfill Finite Element Method
		\item FFT \dotfill Fast Fourier Transformation
		\item FNO \dotfill Fourier Neural Operator
		\item MC \dotfill Monte Carlo
		\item MSE \dotfill Mean Square Error
		\item NN \dotfill Neural Network
		\item PDE \dotfill Partial Differential Equation
		\item PINN \dotfill Physics Informed Neural Network
		\item PINO \dotfill Physics Informed Neural Operator
		\item PPDE \dotfill Parametric Partial Differential Equation
		\item SR \dotfill Squared Residual		
		\item TF \dotfill Transfer Learning
	\end{itemize}

	\begin{acknowledgement}
		The authors cordially thank Mr. Emre Sahin for his contribution to the present work.
	\end{acknowledgement}

	\printbibliography

\end{document}